 \documentclass[12pt]{article}
\usepackage{amsmath,amssymb,bm}
\usepackage[ left=1.5cm,right=1.25cm,top=1.25cm,bottom=1.25cm]{geometry}
\usepackage{times}
\usepackage{setspace}
\usepackage{booktabs}

\RequirePackage{graphicx}

\RequirePackage{natbib}
\usepackage{url}

\usepackage[english]{babel}
\usepackage{array}
\usepackage{rotating}
\usepackage{color}

\usepackage{multirow}
\usepackage{enumerate}
\usepackage{epstopdf}
\usepackage{float}

\usepackage{bbm}
\usepackage{subcaption}
\RequirePackage{amsthm,amsmath,amsfonts,amssymb}
\usepackage{mathrsfs}
\RequirePackage[colorlinks,citecolor=blue,urlcolor=blue]{hyperref}

\usepackage[bottom]{footmisc}
\usepackage{enumitem}
\usepackage{subcaption}

\newcounter{magicrownumbers}

\newtheorem{thm}{Theorem}[section]
\newtheorem{proposition}[thm]{Proposition}

\theoremstyle{definition}

\RequirePackage{geometry}
\RequirePackage{t1enc}
\RequirePackage{enumerate}
\RequirePackage{epsfig}
\RequirePackage{makeidx}
\RequirePackage{graphicx}
\RequirePackage{listings}
\RequirePackage{gensymb}

\usepackage{csvsimple}

\usepackage[utf8]{inputenc}
\usepackage[utf8]{luainputenc}
\usepackage[english]{babel}

\usepackage{xcolor}
\usepackage{caption}

\usepackage{algorithm}
 \usepackage{algpseudocode}

\newtheorem*{restatedproposition}{Proposition 2.1 (restated)}

\title{Prescreen, Then Subsample: Efficient model training for large-scale biomedical data with correlated covariates}

\author{
Subhadra Dasgupta$^{1}$, Christian Staerk$^{1,2}$ \\[0.5em]
{\small $^{1}$ IUF -- Leibniz Research Institute for Environmental Medicine, Düsseldorf, Germany}\\[-0.5em]
{\small $^{2}$ Department of Statistics, TU Dortmund University, Dortmund, Germany}
}

\date{}

\begin{document}

\doublespacing

        \maketitle 


\vspace{-0.5cm}
\begin{center}
    \textbf{Abstract}
\end{center}
\vspace{-0.2cm}

Statistical modeling of high-dimensional and large-sample data with complex correlation structures is increasingly important in biomedical research, particularly in large-scale omics applications such as genomic studies. Analyzing these large-scale datasets imposes substantial computational and memory demands, requiring efficient model training strategies that maintain predictive accuracy. We propose \textit{Prescreen, Then Subsample (PTS)}, a two-stage approach that combines variable selection with optimal-design-based informative subsampling to address both high dimensionality and large sample size. The first stage screens informative variables and can explicitly account for covariate correlation; the second stage selects informative observations in the prescreened variable space. In contrast to existing two-stage subdata selection approaches, PTS can incorporate multivariable fine-mapping to identify a correlation-aware covariate space for informative subsampling. In particular, we integrate the Sum of Single Effects (SuSiE) model for fine-mapping-based prescreening with information-based optimal subdata selection (IBOSS) for sample selection. 
Simulation studies based on correlated Gaussian covariates and biobank-scale synthetic genotype data show that PTS reduces model training time while achieving predictive performance comparable to full-data training. We find that SuSiE-based PTS provides improved predictive performance consistent with better stage-one screening precision compared with commonly used univariate prescreening methods. PTS therefore provides a flexible and scalable training framework for large-scale biomedical data with complex correlation structures, including genomic data with linkage disequilibrium. 
\par\vspace{2pt}

\section{Introduction}

Recent advances in high-throughput technologies and large-scale cohort studies  have enabled the generation of large and high-dimensional datasets. For example, genotype data may contain hundreds of thousands or millions of single-nucleotide polymorphisms~(SNPs) measured across biobank-scale cohorts, providing the basis for polygenic score modeling of a wide range of phenotypes and disease risks~\citep{choi2020tutorial}. Fitting regression models with commonly used shrinkage methods to such data is computationally expensive due to both high dimensionality and large sample sizes~\citep{qian2020fast}. To improve computational efficiency, variable selection and  optimal-design-based informative subsampling provide complementary strategies for reducing the number of covariates and observations used for model training, respectively.

Variable selection addresses high dimensionality by prescreening informative covariates before model training. Common prescreening approaches include Sure Independence Screening (SIS) \citep{fan2008sure} based on marginal association screening, and the directly related Thresholding (TH) approach widely used in genome-wide association studies~(GWAS). In genomic applications, the strong local correlation among genetic variants due to linkage disequilibrium~(LD) motivates methods that explicitly account for the correlation structure among covariates. Clumping and Thresholding~(CT) combines marginal association screening with LD-based clumping to retain approximately independent variants~\citep{euesden2015prsice, choi2020tutorial}. Fine-mapping methods instead jointly analyze correlated variants within genomic regions using multivariable models to distinguish multiple association signals from correlated proxies. One prominent example is the Bayesian Sum of Single Effects~(SuSiE) model for identifying signals in sparse and highly correlated settings~\citep{wang2020simple}.    

In contrast, subsampling methods address large sample sizes  by selecting a smaller but informative subset of observations for model training. Many informative subsampling methods have been developed mainly for low-dimensional settings and are motivated by optimal design principles, aiming to retain as much information as possible in a subsample of small size. Compared with uniform random subsampling, informative subsampling can improve estimation efficiency and predictive performance~\citep{ma2015leveragingJMLR, wang2019information, wang2021orthogonal}. For instance, Information-Based Optimal Subdata Selection (IBOSS) \citep{wang2019information} approximates D-optimal subsampling by selecting observations with extreme covariate values, while other approaches construct subsamples that approximate orthogonal arrays \citep{wang2021orthogonal, ren2021subdata}. Other approaches include ODBSS \citep{dasgupta2025efficient}, which allows the optimality criterion to reflect the estimation objective, and probabilistic subsampling based on leverage scores~\citep{ma2015leveragingJMLR} or A-optimality~\citep{ai2021optimal}.

Only recently has the literature begun to combine variable selection with informative subsampling to address settings with both large numbers of covariates and large sample sizes. SIS-IBOSS \citep{wang2024efficient} first applies SIS to screen a reduced set of covariates and then uses IBOSS for subsampling. The CLASS (Combining LASSO and Subdata Selection) algorithm \citep{singh2023subdata} similarly selects variables through repeated LASSO fits on random subsets before applying IBOSS. Further recent two-stage approaches include FAME~\citep{wei2024efficient}, which combines marginal feature screening with information-based observation reduction, and \citet{chasiotis2026efficient}, who combine LASSO-based variable selection with leverage-score subsampling. However, biomedical applications pose additional challenges, including very high dimensionality, sparsity and complex local correlation structures such as LD. For such high-dimensional data, prescreening should not only have high power but also high precision, because retaining non-informative or redundant variables can direct subsequent informative subsampling toward irrelevant covariate dimensions. In the presence of complex covariate correlation structures, correlation-aware multivariable prescreening can be especially important for identifying the relevant signal structure. 

Motivated by these challenges, we propose Prescreen, Then Subsample~(PTS), a general two-stage framework that combines variable selection with optimal-design-based subsampling for high-dimensional and large-sample data. PTS is designed to reduce model-training time while preserving predictive accuracy. In the first stage, PTS performs prescreening to identify a reduced set of informative variables; in the second stage, it selects informative observations within the prescreened covariate space. Consequently, the quality of prescreening directly influences the efficiency of the subsequent informative subsampling step. The PTS framework allows correlation-aware multivariable prescreening and, in particular, can incorporate fine-mapping-based SuSiE before applying subsampling. This makes PTS particularly suitable for genomic and other biomedical data with strongly correlated covariates. Additionally, the two-stage structure of PTS reduces memory requirements and allows model training using methods that may otherwise be impractical on the full dataset.

The rest of the paper is organized as follows: Section~\ref{Section: Methods} introduces the PTS framework and provides an analytical illustration of how prescreening errors can affect prediction risk and the statistical efficiency of informative subsampling. Section~\ref{Section: Numerical Results} evaluates PTS in simulation studies with correlated covariates, showing that PTS can achieve predictive performance comparable to full-data model training while substantially reducing computation time. Section~\ref{Section: HAPNEST Data Example} illustrates the scalability of PTS for polygenic score modeling using biobank-scale HAPNEST genotype data with more than 100,000 SNPs and 168,000 individuals.

\section{Methods}
\label{Section: Methods}

Let $I=\{1,\ldots,n\}$ and $J=\{1,\ldots,p\}$ denote the index sets of $n$ observations and $p$ covariates, respectively. The full dataset is denoted by $\mathcal{D}_{IJ} = \bigl\{(\bm{x}_{iJ}, y_i) : i \in I\bigr\}$, where \(\bm{x}_{iJ}=(x_{ij})_{j\in J}\in\mathbb{R}^{p}\) is the covariate vector and $y_i \in \mathbb{R}$ is the  response for observation \(i\). The pair $(\bm{x}_{iJ},y_i)$ is a realization of the random vector $(\bm{x}^\top,Y)^\top$, where $\bm x=(X_1,\ldots,X_p)^\top$. We assume that the conditional distribution of $Y$ given $\bm{x}$ belongs to an exponential family, with conditional mean modeled as  
\begin{equation}
\eta(\mathbb{E}[Y \mid \bm{x}]) = \beta_0 + \bm{x}^\top \tilde{ \bm{ \beta} },
\label{eq:linear_model} 
\end{equation}
where $\bm{\beta}=(\beta_0,\ \tilde{ \bm{ \beta} }^\top)^\top\in\mathbb{R}^{p+1}$ denotes the parameter vector and $\eta$ is the link function. The design matrix corresponding to $\mathcal{D}_{IJ}$ is denoted by
$ \bm{X}_{IJ} =
\bigl[\bm{x}_{1J},\ldots,\bm{x}_{nJ}\bigr]^\top
\in\mathbb{R}^{n\times p}$, and the response vector is denoted by $\bm{y}_{I}=(y_1,\ldots,y_n)^\top\in\mathbb{R}^{n}$. 

The computational cost of model training depends on both the sample size~$n$ and the number of covariates~$p$. To understand this dependence, consider the case of a Gaussian response and the identity link, for which equation~\eqref{eq:linear_model} reduces to the standard linear regression model. For ordinary least squares with $n > p$, model fitting typically requires $\mathcal{O}(np^2)$ operations. For LASSO estimation using coordinate descent, in high-dimensional settings with $p > n$, the computational cost for a fixed regularization parameter is approximately  $\mathcal{O}(np s_{iter})$, where $s_{iter}$ denotes the number of complete coordinate-descent iterations required for convergence \citep[][]{friedman2010regularization}. For applications with large values of $n$ or $p$, multivariable model training on the full dataset $\mathcal{D}_{IJ}$ can be computationally expensive or even infeasible and may impose substantial  memory requirements. These challenges are particularly pronounced in polygenic risk score (PRS) modeling, where hundreds of thousands or millions of genetic variants are analyzed in biobank-scale cohorts as shown in \cite{choi2019prsice,prive2018efficient}.  

A common approach to address high dimensionality, that is, a large number of covariates~\(p\), is to use variable selection techniques. A subset $J'\subset J$ of informative covariates ($|J'| = q \ll p$) is identified, and the reduced dataset $\mathcal{D}_{IJ'}$ is used for model training. Subsampling methods, on the other hand, address large sample sizes~$n$ by selecting a reduced set of informative observations. A subset $I'\subset I$ of informative observations ($|I'| = k \ll n$) is identified, and the reduced dataset $\mathcal{D}_{I'J}$ is used for model training. Both variable selection and subsampling reduce model-training time and memory requirements by reducing the number of covariates and observations, respectively.

\begin{figure}[t]
	\centering
		\includegraphics[width=\linewidth]{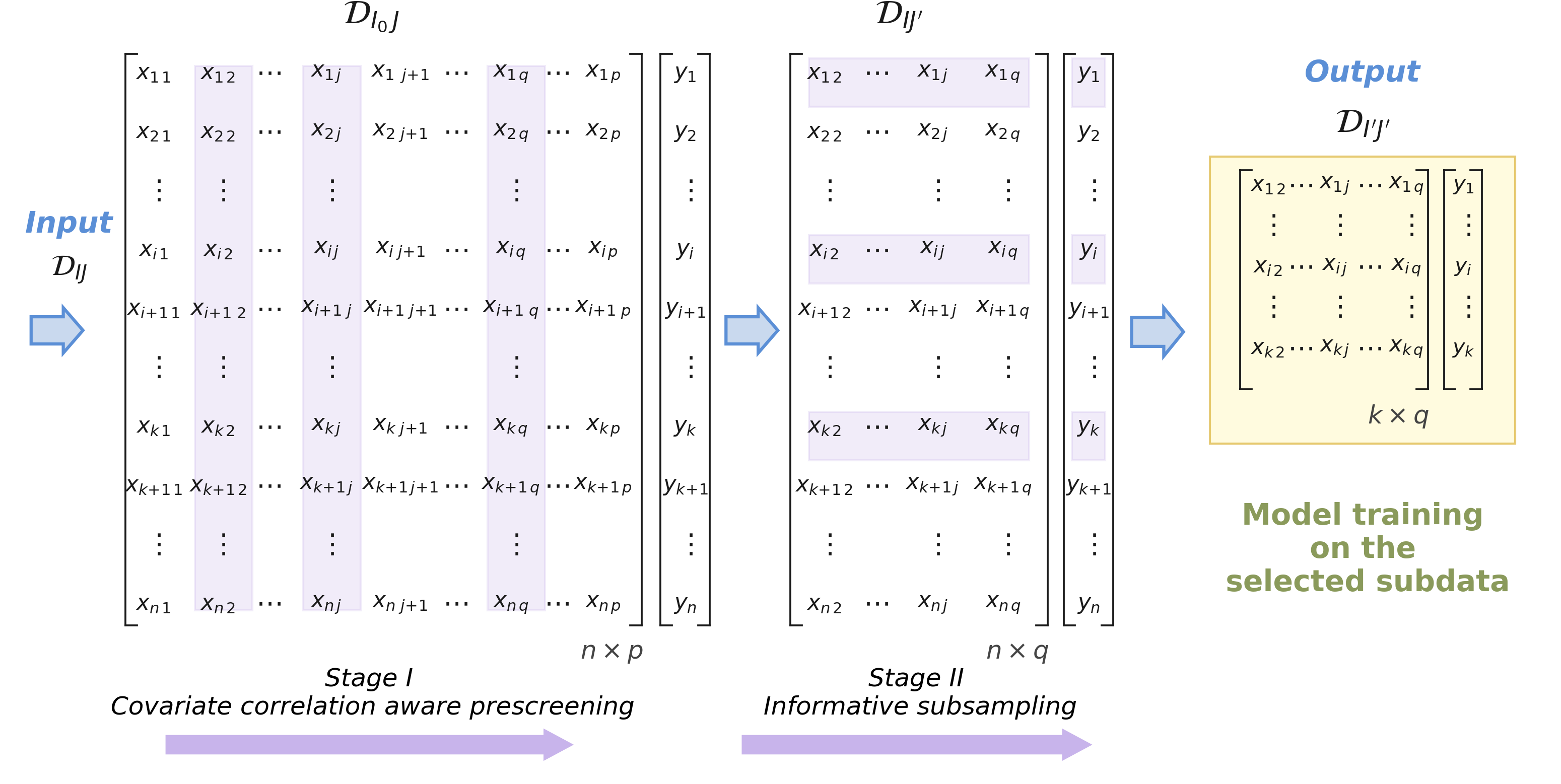}
	   	\caption{ \it Schematic representation of PTS for the case $c =1$.}
	\label{Fig: Graphical Abstract} 
\end{figure}

This paper focuses on scalable modeling for high-dimensional, large-sample data and targets biomedical applications characterized by correlated covariate structures. To address these challenges, we propose \textit{Prescreen, Then Subsample~(PTS)}, a general two-stage subdata selection framework that combines prescreening and informative subsampling for scalable model training, as described in Algorithm~\ref{alg:subsample}. PTS first identifies a reduced outcome-relevant covariate space ($\mathcal{D}_{IJ'}$) using a random sample of the full data denoted by $\mathcal{D}_{I_0J}$, where $|I_0|=\lfloor cn\rfloor$ and $c\in(0,1]$, and then selects informative observations within this space ($\mathcal{D}_{I'J'}$). 
By restricting observation selection to the screened covariates, PTS aims to concentrate the subsequent subsampling effort on dimensions relevant for prediction. Figure~\ref{Fig: Graphical Abstract} provides a schematic representation of PTS. Details of Stage~I (prescreening) and Stage~II (informative subsampling) are provided in Section \ref{subsection: stage I  description} and Section \ref{subsection: stage II  description}.

\begin{algorithm}[t]
\caption{Prescreen, Then Subsample (PTS)}\label{alg:subsample}
\begin{algorithmic}[1]
\Statex \textbf{Input:} Dataset $\mathcal{D}_{IJ}$ with $|I| = n$ observations and $|J| = p$ covariates, prescreening sample proportion~$c\in(0,1]$, target subsample size $k \ll n$, prescreening method, informative subsampling method, and prediction method

\Statex \textbf{Aim:} Reduce model training time while maintaining prediction accuracy

\State \textbf{Stage I: Covariate prescreening}
\Statex I.a) Select a random subset of observations $I_0\subseteq I$ without replacement, with  $|I_0|=\lfloor cn\rfloor$, yielding the screening dataset $\mathcal{D}_{I_0J}$
\Statex I.b) Apply the prescreening method to $\mathcal{D}_{I_0J}$ to identify a reduced covariate set
$J' \subseteq J$, with $|J'| = q \ll p$

\State \textbf{Stage II: Informative subsampling}
\Statex II.a) Apply the optimal-design-based informative subsampling method to $\mathcal{D}_{IJ'}$ to select a subset of observations $I' \subseteq I$ with $|I'| = k \ll n$ 

\Statex \textbf{Final reduced dataset:} $\mathcal{D}_{I'J'}$ with $k$ observations and $q$ covariates  

\Statex \textbf{Model fitting:} Fit the prediction model on $\mathcal{D}_{I'J'}$ using the specified prediction method
\Statex \textbf{Output:} Reduced dataset $\mathcal{D}_{I'J'}$ and fitted prediction model

\end{algorithmic}
\end{algorithm}

\subsection{Stage I: Covariate prescreening} \label{subsection: stage I  description}

High-dimensional modeling with large $p$ is often characterized by sparsity, where only a small subset of variables contains signal, while the remaining variables are noise. Training models on all $p$ variables can therefore be inefficient and computationally expensive. In Stage I, PTS prescreens a small set of informative variables using a variable selection method. 

In this paper, we first consider covariate prescreening using two closely related univariable methods, sure independence screening (SIS) proposed by \cite{fan2008sure} and thresholding (TH)~\citep{choi2020tutorial,euesden2015prsice}, which are commonly used in biomedical applications. For high-dimensional linear regression modeling, SIS ranks predictors according to their marginal correlation with the response, $|\text{corr}(\boldsymbol{x}_j, \bm{y})|$. It then selects the top $q$ variables for further modeling. Under assumptions of sufficiently strong signals and controlled covariate collinearity, SIS has the \emph{sure screening property}, meaning that all truly relevant variables are retained with a high probability. TH is widely used in genome-wide association studies (GWAS). TH fits a marginal model for each covariate and retains variables whose associated $p$-values pass a predefined threshold. In GWAS, a common Bonferroni-corrected threshold $\alpha/p$ ($\alpha= 0.05 $) is used  to control the family-wise error rate across the $p$ tests. 

However, univariable methods do not account for the correlation structure among covariates, which is particularly relevant in genomic data due to linkage disequilibrium (LD).  Therefore, correlation-informed variable selection procedures are needed to reduce redundancy among highly correlated covariates. We consider thresholding combined with clumping (CT), which addresses local correlation by retaining strongly associated and approximately independent covariates~\citep{choi2020tutorial,euesden2015prsice}. Starting from variables identified by TH, clumping retains the variant with the strongest marginal association within a local covariate correlation region (LD region) and removes highly correlated variants in the same region. However, a potential pitfall is that a correlated region may contain more than one distinct association signal, so that clumping may discard additional informative variants.

 In contrast to marginal screening and clumping-based procedures, fine-mapping methods jointly analyze correlated variants within a region and aim to separate distinct association signals from correlated proxies. We also consider the Sum of Single Effects~(SuSiE) approach, which is a Bayesian fine-mapping method for sparse settings that represents the joint effect vector as a sum of $L$ single-effect components, where $L$ is an upper bound on the number of distinct association signals~\citep{wang2020simple}. SuSiE uses a variational approximation to the posterior distribution to estimate posterior inclusion probabilities~(PIPs), representing the posterior probability that the variable contributes to at least one association signal. The final PIP for variable $j$ is $\mathrm{PIP}_j = 1 - \prod_{\ell=1}^{L}(1-\gamma_{\ell j})$, where $\gamma_{\ell j}$ is the posterior probability that variable $j$ is selected in component $\ell$. Thus, variables with larger PIPs are considered to have stronger evidence of association. Further, when the covariates are highly correlated, their association signals may be difficult to distinguish, resulting in low individual PIPs (as the posterior probability is distributed among them). SuSiE additionally provides 95\% credible sets $\mathcal{C}_\ell \subseteq J$ for some of the $L$ single-effect components. Each credible set is the smallest set of variables whose cumulative component-specific posterior probabilities $\gamma_{\ell j}$ reach at least 95\%. Thus, under the variational posterior approximation, $\mathcal{C}_\ell$ contains the variable selected by the corresponding single-effect component with at least 95\% posterior probability.

 For the numerical study and polygenic risk modeling application in Sections~\ref{Section: Numerical Results} and \ref{Section: HAPNEST Data Example}, we consider TH, CT and SuSiE as alternative prescreening methods for selecting the reduced covariate set $J'$. The resulting dataset $\mathcal{D}_{IJ'}$ is used in Stage~II of PTS for informative subsampling across all $n$ observations. More specific implementation choices are described in Section~\ref{Label_Section_Simulation_Setup}.

\subsection{Stage II: Informative subsampling} \label{subsection: stage II  description}

Even after reducing the number of covariates from $p$ to $q$ in Stage~I, model training using all observations can remain computationally expensive when the sample size~$n$ is large. In such cases, rather than using the entire prescreened dataset~$\mathcal{D}_{IJ'}$, a subset $I'\subset I$ of $k\ll n$ informative observations can be selected, yielding the reduced dataset~$\mathcal{D}_{I'J'}$ for model training.

Informative subsampling methods have been developed for low-dimensional regression settings, particularly when $q\ll n$, and are often motivated by optimal design principles aimed at improving estimation efficiency. Previous studies have shown that informative subsampling can improve estimation and prediction performance compared to uniform random subsampling of the same size \citep{ma2015leveragingJMLR, allen2017near, LogisticWang2018optimal, wang2019information, wang2021orthogonal, ren2021subdata, dasgupta2025efficient}. Furthermore, in many cases, informative subsampling methods have been observed to achieve prediction and estimation errors comparable to those obtained using the full dataset, despite relying on a relatively small subset of informative observations. A common approach to selecting such informative observations is based on optimality criteria defined in terms of the Fisher information matrix (FIM).

For a parametric regression model, the Fisher information matrix (FIM) quantifies the available information for estimating the regression parameter~$\bm{\beta}_{J'}$. Here, $\bm{\beta}_{J'}$ denotes the $(q+1)$-dimensional parameter vector of the regression model fitted on the prescreened covariates $J'$. For the regression model in equation~\eqref{eq:linear_model} restricted to the prescreened covariates~$J'$ and with conditionally independent observations, the FIM based on the prescreened dataset $\mathcal{D}_{IJ'}$  is given by 
\[\mathcal{I}{(\bm{\beta}_{J'}, \mathcal{D}_{IJ'})} =\sum_{i\in I}     \mathcal{I}{(\bm{\beta}_{J'}, \bm{x}_{iJ'})},\] 
where the information contribution of observation~$i$ is  
\[\mathcal{I}{(\bm{\beta}_{J'}, \bm{x}_{iJ'})} = \mathbb{E}_{\bm{\beta}_{J'}}
\left[
\left\{
\frac{\partial \ell(y_i\,|\,\bm{\beta}_{J'}, \bm{x}_{iJ'})}
{\partial\bm{\beta}_{J'}}
\right\}^{\!\top}
\left\{
\frac{\partial \ell(y_i\,|\,\bm{\beta}_{J'}, \bm{x}_{iJ'})}
{\partial\bm{\beta}_{J'}}
\right\}
\right],\]
and $\ell(y_i\,|\,\bm{\beta}_{J'},\bm{x}_{iJ'})$ is the log-likelihood of observation~$i$. Here, derivatives of the log-likelihood with respect to the regression parameters are treated as row vectors. 
Let $\bm{\delta} \in \{0,1\}^n $ denote a subsample selection vector, where $ \delta_i = 1 $ indicates that the $i$-th observation is included in the subsample~$\mathcal{D}_{I'J'}$, that is, $I'=\{i\in I:\delta_i=1\}$ and $\sum_{i=1}^n  \delta_i = k  $. The FIM corresponding to the selected subsample is given by 
$\mathcal{I}{(\bm{\beta}_{J'}, \mathcal{D}_{I'J'})} = \sum_{i\in I}    \delta_i   \mathcal{I}{(\bm{\beta}_{J'}, \bm{x}_{iJ'})}$. The optimal subsample of size $k$ is then determined by
\begin{equation}
\bm{\delta}_{\mathrm{opt}}
=
\arg\max_{\bm{\delta}\in\{0,1\}^n}
\Phi\!\left(\mathcal{I}(\bm{\beta}_{J'}, \mathcal{D}_{I'J'} )\right), 
\text{ subject to } \sum_{i=1}^n \delta_i = k,
\label{eq:optimal_subsample_general}
\end{equation}
where $\Phi$ denotes a concave scalar optimality criterion used to quantify the information contained in the selected subsample~\citep{Puke1993}. For a positive definite information matrix $\bm M$, common optimality criteria include D-optimality, corresponding to $\Phi(\bm M)=\log\det(\bm M)$, which aims to minimize the volume of the confidence ellipsoid for the regression parameters, and A-optimality, corresponding to $\Phi(\bm M)=\{\mathrm{tr}(\bm M^{-1})\}^{-1}$, which aims to minimize the sum of the variances of the parameter estimators. 

Since obtaining the exact optimal subsample is generally computationally expensive, informative subsampling algorithms employ approximations to achieve computational efficiency. For example, Information-Based Optimal Subdata Selection (IBOSS), an extreme-value-based subsampling algorithm proposed by \cite{wang2019information} for linear regression models, approximates D-optimal subsampling by selecting observations with extreme marginal covariate values. In this paper, we employ IBOSS for informative subsampling in Stage~II of the proposed PTS framework. Specifically, IBOSS is applied to the prescreened dataset~$\mathcal{D}_{IJ'}$ containing $q$ covariates and selects observations with extreme values of the prescreened covariates. In our implementation, we sequentially select $\lceil k/(2q)\rceil$ observations from each marginal tail, excluding observations selected for preceding covariates and removing duplicates. If more than $k$ observations are selected, we randomly retain $k$ observations.

\subsection{Combining prescreening and informative subsampling}

Model training with high-dimensional, large-scale data can substantially benefit from selecting a small set of informative variables and a small number of informative observations. Specifically, these reductions can improve computational efficiency without compromising prediction accuracy. In high-dimensional settings with many noise variables, informative subsampling without prior variable prescreening may allocate design effort along non-informative directions, thereby reducing the statistical efficiency of the selected subsample. Identifying an informative covariate space~$J'\subset J$ with $|J'|=q$ before subsampling can therefore help to direct subsampling effort toward dimensions that are relevant for prediction.

In PTS, the two stages are directly linked as informative subsampling in Stage~II is performed within the reduced covariate space~$J'$ after prescreening in Stage~I. Two types of prescreening errors can affect subsequent subsampling: if relevant variables are omitted, these informative directions are no longer available to guide subsampling; on the other hand, if many non-informative or redundant variables are retained, part of the available subsampling effort is directed toward less informative dimensions. 

In the presence of strongly correlated covariates, marginal screening methods such as SIS and Thresholding may select redundant proxy variables that are strongly correlated with relevant variables, while relevant variables with weaker marginal associations may be missed. Furthermore, several distinct signals can occur within the same correlated region. As a result, marginal screening methods may select a covariate space for subsampling in Stage~II that does not adequately represent the underlying multivariable signal structure. Multivariable selection methods aim to address these limitations of marginal screening by identifying relevant variables jointly while accounting for the correlation structure among covariates. In genomic applications, fine-mapping methods such as SuSiE~\citep{wang2020simple} can provide a more informative covariate space for subsequent subsampling. Importantly, the power and precision of prescreening can affect both the information available for prediction and the covariate directions used to guide informative subsampling in Stage~II. 

\subsubsection*{Effects of prescreening quality on prediction and informative subsampling}

In Proposition~\ref{Propsition PTS}, we illustrate two consequences of prescreening errors: false-negative screening can increase prediction risk, whereas false-positive screening can reduce the statistical efficiency of informative subsampling under an IBOSS-inspired population-tail design. We consider a simple Gaussian linear regression setting. Let $X_1,X_2,W\sim N(0,1)$  and $\varepsilon\sim N(0,\sigma^2)$ be independent with $\sigma^2>0$. For $\beta_1,\beta_2\neq 0$ and $0<\rho<1$, let
\begin{equation}
Y=\beta_1X_1+\beta_2X_2+\varepsilon,
\qquad
X_3=\rho X_2+\sqrt{1-\rho^2}W.
\label{eq:example}
\end{equation}
Thus, $X_1$ and $X_2$ are the relevant predictors, while $X_3$ is a correlated proxy for $X_2$ with no additional effect on $Y$.

 \begin{proposition}[]\label{Propsition PTS}Under the Gaussian regression setting in \eqref{eq:example},
the following results hold:
\vspace{-2mm}
\begin{enumerate}
\item[(i)] \textbf{False-negative prescreening:} Suppose that $0<|\beta_1|<\rho|\beta_2|$, so that selecting the two covariates with the largest absolute population marginal correlations yields  $\widetilde J=\{2,3\}$ instead of
$J'=\{1,2\}$.  Define the minimum squared linear prediction risk as
$R(J)=\inf_{\bm\alpha}
E\bigl[(Y-{\textstyle\sum_{j\in J}}\alpha_jX_j)^2\bigr]$. 
Then,
\begin{equation*}
R(J')=\sigma^2,\qquad
R(\widetilde J)=\sigma^2+\beta_1^2,\qquad
R(\widetilde J)-R(J')=\beta_1^2>0.
\end{equation*}
Thus, omitting $X_1$ increases the minimum squared linear prediction risk by $\beta_1^2$.

\item[(ii)] \textbf{False-positive prescreening:} Suppose one prescreening method selects $J'=\{1,2\}$, whereas another additionally retains the redundant variable~$X_3$ and selects $J''=\{1,2,3\}$. For a common fixed threshold $t>0$, consider an IBOSS-inspired population-tail design defined by
$\xi_{J,t}(\bm{x})=\dfrac{1}{|J|}\sum_{j\in J}f(\bm{x}\mid |X_j|>t)$, where $f(\bm{x})$ denotes the joint population density of $\bm{x}=(X_1,X_2,X_3)^\top$.
Assuming an oracle final model fitted on the true active
variables $\bm{z}=(X_1,X_2)^\top$, let 
$M_J(t)=E_{\xi_{J,t}}
(\bm{z}\bm{z}^\top)$ denote the normalized Fisher information matrix (up to the common factor $1/\sigma^2$). For the A- and D-optimality criteria $\Phi_A(M)=\{\operatorname{tr}(M^{-1})\}^{-1}$ and
$\Phi_D(M)=\log\det(M)$, respectively, we obtain 
\[
\Phi_A(M_{J'}(t))>\Phi_A(M_{J''}(t)), \qquad
\Phi_D(M_{J'}(t))> \Phi_D(M_{J''}(t)).
\] 
Thus, under this population-tail design, including the redundant variable yields a strictly less informative design according to both the A- and D-optimality criteria.
\end{enumerate}
\end{proposition}

The population-tail construction in Proposition~\ref{Propsition PTS} is motivated by continuous subsampling designs in Corollary 3.2 of \cite{GlemserSchwabe2026}, for which population-tail selection provides an approximate analogue of IBOSS in one dimension. Here, $\xi_{J,t}$ is our coordinate-wise extension and mimics the equal allocation of subsampling effort across the coordinates in $J$, while replacing finite-sample order-statistic selection by the corresponding population tails. Thereafter, we use the normalized information matrix \(M_J(t)=E_{\xi_{J,t}}(\bm{z}\bm{z}^\top)\), as is standard in approximate design theory \citep[see][]{Puke1993}, to compare the information provided by the resulting subsampling designs for estimating \(\beta_1\) and \(\beta_2\) under different prescreening sets. By assuming an oracle final model fitted on the true active variables, we isolate the effect of prescreening on the subsequent subsampling step. In particular, the efficiency loss arises because the redundant variable $X_3$ influences which observations are selected, even though it is not included in the final regression model.

The two parts of Proposition~\ref{Propsition PTS} illustrate the importance of power and precision in prescreening: omitting relevant variables can increase prediction risk, while retaining redundant variables can reduce the statistical efficiency of informative subsampling. The result in part (ii) is exact for the specified population-tail design at a common fixed threshold, but does not establish a general efficiency ordering for finite-sample IBOSS. The detailed proofs of Proposition~\ref{Propsition PTS} are in Supplementary Section S.5.

\section{Numerical Study} \label{Label_Section_Simulation_Setup}
\label{Section: Numerical Results}

This section uses simulation studies to evaluate the performance of  PTS (Algorithm~\ref{alg:subsample}) for high-dimensional, large-scale linear regression when covariates are highly correlated. The quality of model training is evaluated based on predictive performance on independent test data  $\mathcal{D}_{I_{\mathrm{test}}J} = \bigl\{(\bm{x}^\mathrm{test}_{iJ}, y^\mathrm{test}_i) : i \in  I_\mathrm{test}  \bigr\}$ of size $|I_\mathrm{test}| = n_\mathrm{test}$ and $  \bm{x}^\mathrm{test}_{i J}  \in\mathbb{R}^{p}$, using the average squared correlation
\begin{equation}
\label{det21} 
 \overline{r}^2_\mathrm{test} =  \dfrac{1}{n_\mathrm{sims}} \sum_{\mathrm{sim} = 1}^{n_\mathrm{sims}} \big [ \operatorname{Cor}^2({\bm y}^{\mathrm{test}(\mathrm{sim})}, {\hat{\bm y}}^{\mathrm{test}(\mathrm{sim})}) \big ],
\end{equation}
where $ \bm{y}^\mathrm{test}=(y_1^\mathrm{test},\ldots,y^\mathrm{test}_{n_\mathrm{test}})^\top$ and $ \hat{\bm{y}}^\mathrm{test} = [\boldsymbol{1} \, \bm{X}^\mathrm{test}_{{I_\mathrm{test}J'}} ]  ~\hat{\boldsymbol{\beta}}_{\mathcal{D}_{I'J'}} $. The computational efficiency of the PTS framework is evaluated based on the average total runtime (in seconds),
\begin{equation}
\overline{\mathcal{T}} =  \dfrac{1}{n_\mathrm{sims}} \sum_{\mathrm{sim} = 1}^{n_\mathrm{sims}} \mathcal{T}(\mathcal{D}_{IJ}^{(\mathrm{sim})}),
\end{equation}
where $\mathcal{T}(\mathcal{D}_{IJ}^{(\mathrm{sim})})$ denotes the total runtime of PTS on dataset~$\mathcal{D}_{IJ}^{(\mathrm{sim})}$; it is the sum of the time taken for prescreening, subsampling, and then model training. Additionally, final variable selection performance is evaluated as a secondary criterion using
$\text{Power} = |\widehat{\mathcal A}\cap\mathcal A|/|\mathcal A|$
and
$\text{Precision} = |\widehat{\mathcal A}\cap\mathcal A|/|\widehat{\mathcal A}|$,
where $\mathcal A=\{j\in\{1,\ldots,p\}: \tilde{\beta}_j \neq 0\}$ denotes the true active set, and $\widehat{\mathcal A}$ denotes the active set selected by the fitted model. 

\subsection{Simulation setup}
We generate the covariates as $\boldsymbol{x}\sim\mathcal{N}_p(\boldsymbol{0},\boldsymbol{\Sigma})$, with $B$ independent blocks of equal size $m$, such that $p=mB$ and $\boldsymbol{\Sigma}=\operatorname{diag}(\boldsymbol{\Sigma}_1,\ldots,\boldsymbol{\Sigma}_B)$. Within each block, we assume an AR(1) correlation structure, $(\boldsymbol{\Sigma}_b)_{ij}=\rho^{|i-j|}$, $\rho\in(-1,1)$, for $i,j=1,\ldots,m$, and $b=1,\ldots,B$, providing a simple approximation to local LD. We set $B=5$ and $\beta_0=1$, and assume a sparse coefficient vector $\tilde{\boldsymbol{\beta}}$ with the following two signal configurations:
\begin{enumerate}
\item \textbf{Fixed 12 signal loci}:
 The number of active signals is fixed at $|\mathcal A|=12$, with $5$, $4$, $2$, $1$, and $0$ signals in blocks $1$--$5$, respectively. Specifically,
$(\tilde{\beta}_{j_1},\ldots,\tilde{\beta}_{j_{12}})
=(1.1,\,1.2,\,0.8,\,0.5,\,2,\,-1.1,\,1.2,\,2,\,1,\,1,\,0.2,\,2)$,
where
$(j_1,\ldots,j_{12})
=(1,\,2,\,\lfloor m/2\rfloor,\,\lfloor m/2\rfloor+1,\,m,\,
m+4,\,m+5,\,m+\lfloor m/2\rfloor,\,m+\lfloor m/2\rfloor+1,\,
2m+1,\,2m+\lfloor m/2\rfloor+1,\,3m+\lfloor m/2\rfloor)$,
and $\tilde{\beta}_j=0$ otherwise.

\item \textbf{Random signal loci}: The sparsity ($s$) is fixed and the active set $\mathcal{A}$ (of size $|\mathcal A|=\lfloor sp \rfloor$) is selected at random in each simulation. For $j\in\mathcal{A}$,
$\tilde{\beta}_j\stackrel{\mathrm{i.i.d.}}{\sim}\mathrm{Rademacher}(1/2)$, that is, $
\Pr(\tilde{\beta}_{j}=1)=\Pr(\tilde{\beta}_{j}=-1)=\tfrac12,
$ 
while $\tilde{\beta}_j=0$ for $j\notin\mathcal{A}$.
\end{enumerate}

The conditional distribution of the response~$Y$ in \eqref{eq:linear_model} is Gaussian with an identity link and $\operatorname{Var}(Y\mid\boldsymbol{x})=\sigma^2$, leading to a linear regression model. The signal-to-noise ratio (SNR) is set to 0.5 by setting $\sigma^2 = \frac{ \tilde{\bm{\beta}}^T  \boldsymbol{\Sigma} \tilde{\bm{\beta}}}{SNR}$ \citep{hastie2020best}. All the results are based on $n_{\mathrm{sims}}=100$ simulation runs. Across the \(n_{\mathrm{sims}}=100\) simulation replicates, new training and test data are generated, including new covariates and response errors. In the `Fixed 12 signal loci' setting, the signal locations and coefficients remain fixed, whereas in the `Random signal loci' setting, both the active set \(\mathcal A\) and coefficient signs are regenerated in each replicate. 

\subsection{PTS implementations}
In the simulation study, three instances of PTS are used with TH, CT, and SuSiE in Stage I, IBOSS in Stage II, and LASSO for model training on the reduced informative dataset. An additional PTS instance employing oracle variable selection in Stage I, based on the known true signal variables, is included as a benchmark to assess the impact of prescreening on overall prediction performance. This oracle selection is followed by IBOSS in Stage II and LASSO for model training. PTS is benchmarked against LASSO models trained on uniformly random subsamples of the same size (\(k\)) as the PTS subsamples, as well as against a LASSO model trained on the full dataset. Except for the \texttt{snpnet} benchmark in Section~\ref{Section: HAPNEST Data Example}, LASSO is fitted using \texttt{cv.glmnet} in \texttt{R}. The regularization parameter ($\lambda$) is selected using the built-in 10-fold cross-validation procedure. Specifically, we use \texttt{lambda.min} returned by \texttt{cv.glmnet}, which is the value of  $\lambda$ that minimizes the mean cross-validation error.

Since \(p=5000\) is substantially smaller than the number of covariates in the biomedical application, implementing all three prescreening procedures, TH, CT, and SuSiE, in \texttt{R} is computationally feasible, and we use these implementations in this simulation study. The TH prescreening retains the covariates when the marginal model yields a $p$-value below $0.05/p$ (Bonferroni-adjusted threshold). The TH step is conceptually similar to the SIS procedure in \cite{wang2024efficient}. In the considered settings, we use TH for two reasons: first, it facilitates the implementation of CT; and second, it provides a simple threshold-based rule for determining which variables are retained. The CT-based screening uses the variables retained by TH and selects the strongest approximately independent representatives using an absolute correlation cutoff of $0.1$ for clumping. Finally, SuSiE is a fine-mapping method for identifying signal variables in settings with correlated covariates. SuSiE is applied to each correlated block containing at least one covariate that passes the Bonferroni-adjusted threshold in the TH step. Within each selected block, SuSiE is fitted using covariates with marginal \(p\)-values below 0.05, following the prefiltering strategy of \cite{lapierre2021identifying}. We use the \texttt{susieR} implementation with $L=10$, which allows up to ten single effects per block, and retain covariates with PIPs greater than 0.5.

\subsection{Simulation Results} \label{SubSection: Simulation results}

In this section, we consider a scenario with $n=100{,}000$ training observations, $n_{\mathrm{test}}=30{,}000$ test observations, $p=5000$, $\rho = 0.95$, and $SNR = 0.5$, for both the `Fixed 12 signal loci' and `Random signal loci' settings. The `Fixed 12 signal loci' setting has $0.24\%$ sparsity, corresponding to $12$ signals, while for the `Random signal loci', we set the sparsity to $s=1\%$, corresponding to 50 signals. We set $c = 0.5$ as a compromise between prescreening performance and computational cost (see Supplementary Figure S.1). For each PTS instance and the uniform subsampling benchmark, we set $k = 625, 1250, 2500, \text{ and } 5000$.

\begin{figure}[t]
\centering	

\begin{minipage}{0.49\textwidth}
    \centering
    \text{\footnotesize Fixed 12 signal loci }
\end{minipage}
\hfill
\begin{minipage}{0.49\textwidth}
    \centering
    \text{\footnotesize Random 50 signal loci }
\end{minipage}
\vspace{0.0em}
	\begin{subfigure}{.50\textwidth}
		\centering
         \includegraphics[width=\linewidth, height=0.3\textheight, keepaspectratio]{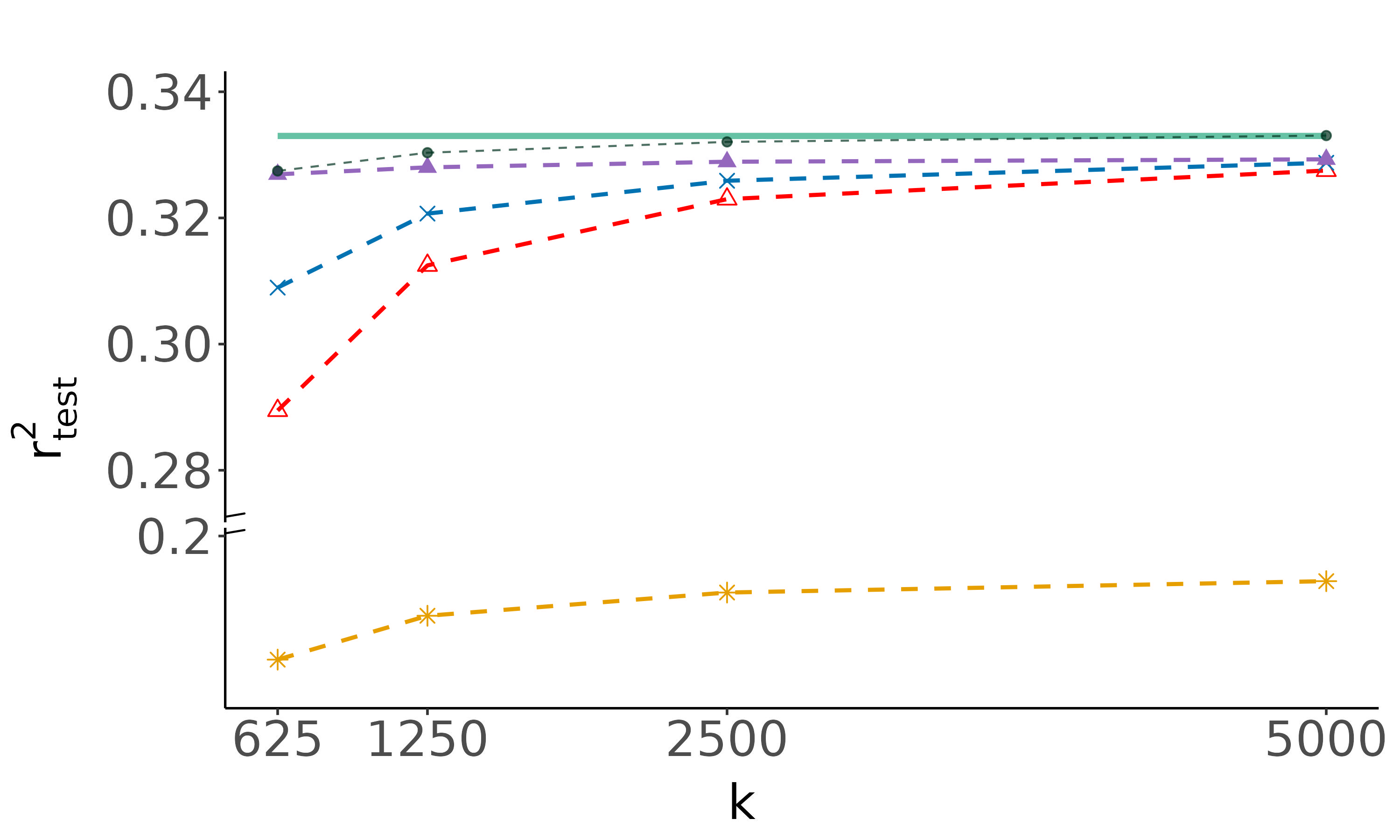}
		
	\end{subfigure}
      \hspace{-0.5em}
    \begin{subfigure}{.50\textwidth}
		\centering
		\includegraphics[width=\linewidth, height=0.3\textheight, keepaspectratio]{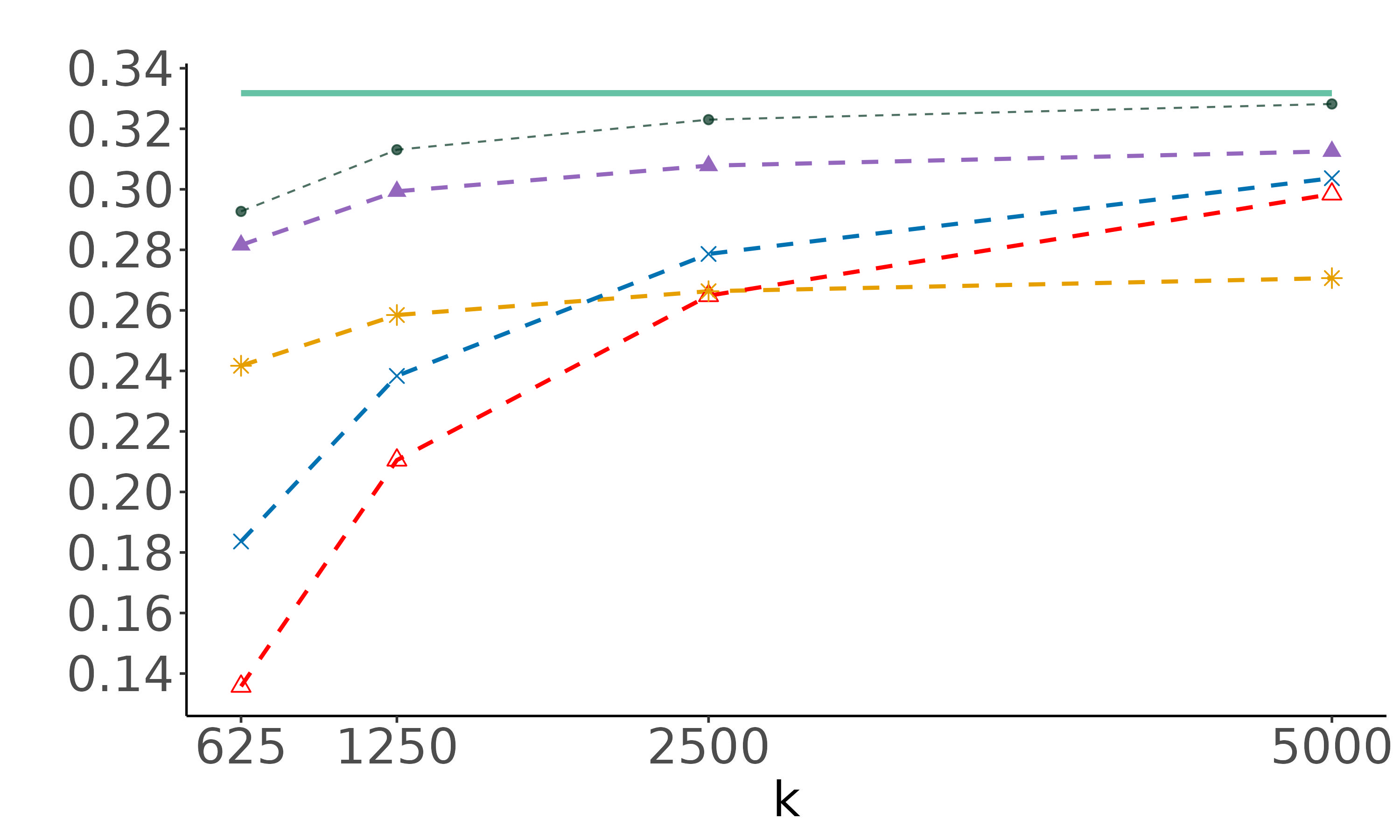}

	\end{subfigure}
{ \it \text{(a) Predictive performance using average $r^2$ on test data}  }	

\vspace{1.0em}

\centering
\begin{minipage}{0.49\textwidth}
    \centering
    \text{\footnotesize Fixed 12 signal loci }
\end{minipage}
\hfill
\begin{minipage}{0.49\textwidth}
    \centering
    \text{\footnotesize Random 50 signal loci }
\end{minipage}
\vspace{0.0em}
	\begin{subfigure}{.50\textwidth}
		\centering
        \includegraphics[width=\linewidth, height=0.3\textheight, keepaspectratio]{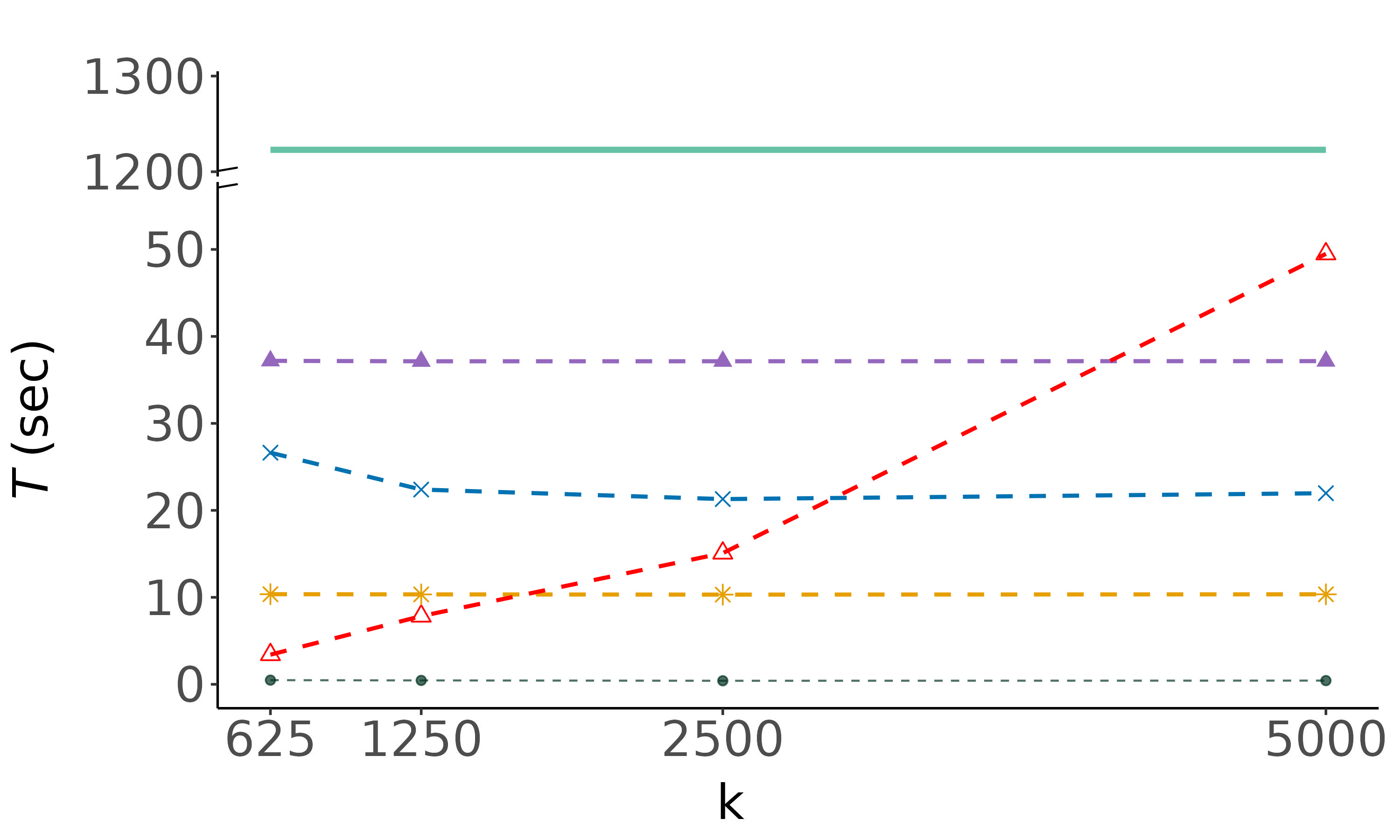}
             
	\end{subfigure}%
     \hspace{-0.5em}
\begin{subfigure}{.50\textwidth}
		\centering
		\includegraphics[width=\linewidth, height=0.3\textheight, keepaspectratio]{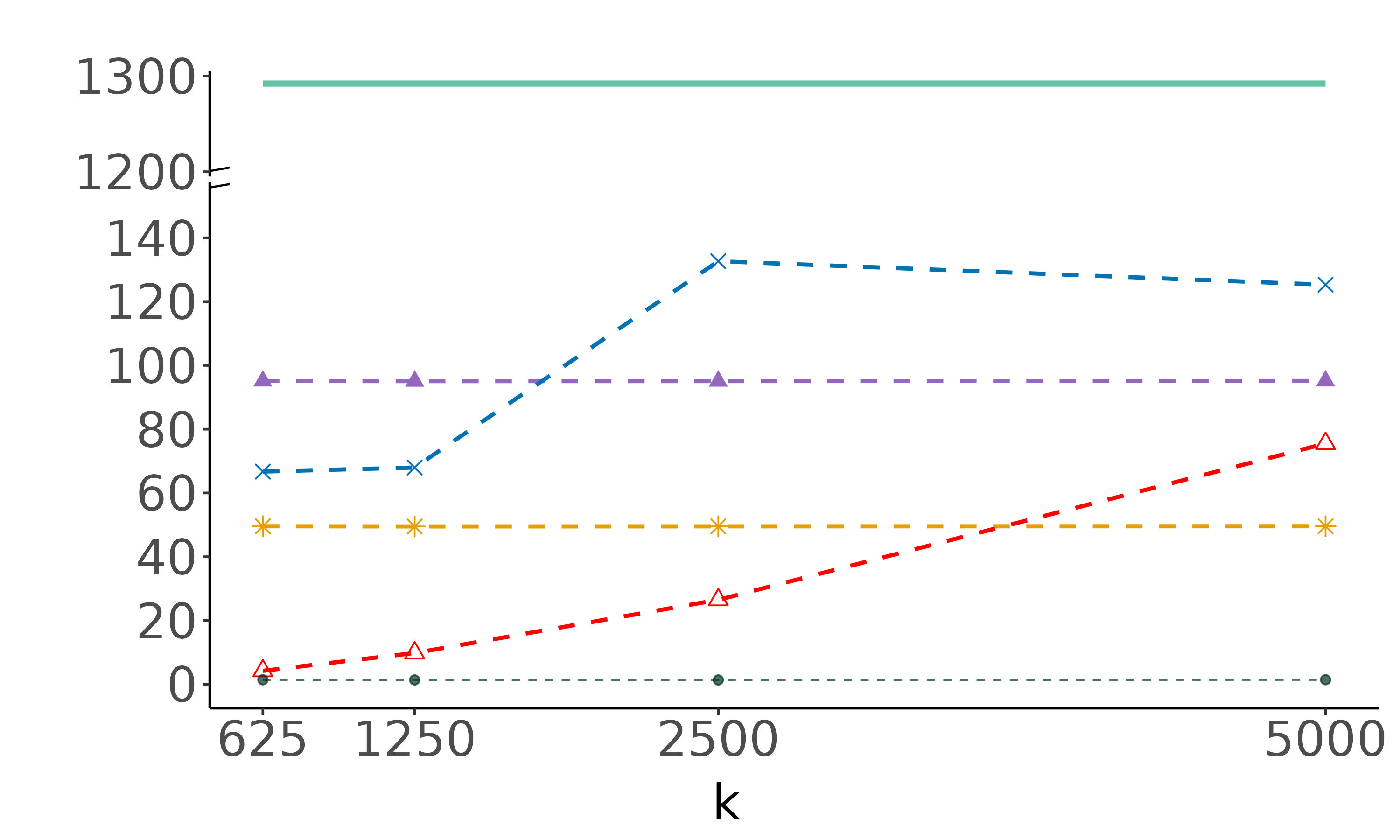}
	\end{subfigure}
      \hspace{-0.5em}
	
  { \it \text{(b) Computational efficiency based on total runtime in seconds}}
\vspace{1.0em} 

   \includegraphics[width=0.95\textwidth]{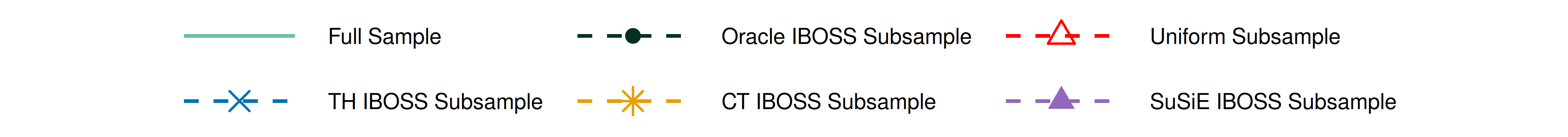} 

\vspace{0.0em}

\caption{\it Benchmarking scalability of PTS based on $100$ simulation runs
for the linear regression model with $\rho = 0.95$, $SNR = 0.5$,
$n = 10^5$, $n_{\mathrm{test}}=30{,}000$, $p = 5000$, and $c = 0.5$.}
\label{figure:instance-comparisons}
\end{figure}

Figure~\ref{figure:instance-comparisons} benchmarks the PTS instances based on predictive performance and total runtime. In both settings, `Fixed 12 signal loci' and  `Random signal loci', SuSiE-IBOSS and TH-IBOSS achieve higher predictive performance than uniform random sampling. In particular, SuSiE-IBOSS attains predictive performance comparable to the full-data model for subsample sizes as small as $k= 2500$ and $5000$ ($2.5\%$ and $5.0\%$ of the full sample size, respectively). SuSiE-IBOSS performs better than TH-IBOSS at smaller subsample sizes, while their predictive performance is similar at $k=5000$. The improved performance of SuSiE-IBOSS is consistent with its higher prescreening precision in Stage I compared with TH (see Supplementary Figure~S.1). Further, the estimated mean test $r^2$ values have narrow $95\%$ confidence intervals (see Supplementary Table~S.1).  At the same time, SuSiE-IBOSS runs approximately 33 and 14 times faster than full-data LASSO training in the `Fixed 12 signal loci' and `Random signal loci' settings, respectively, while TH-IBOSS achieves speedups of approximately 56 and 10 times, respectively, for a subsample size of \(k=5000\). At \(k=5000\), the total runtime of PTS and uniform subsampling is comparable. For small subsample sizes (e.g., \(k=625\)), PTS can be slower than uniform subsampling because of the additional prescreening and subdata-selection steps. However, the total runtime of SuSiE-IBOSS and TH-IBOSS remains nearly constant across \(k\), since model training is performed in the reduced prescreened covariate space, whereas the runtime under uniform subsampling increases sharply with \(k\) because the full \(p\)-dimensional feature set is retained. This demonstrates the scalability of PTS to high-dimensional, large-scale data and highlights the particular advantage of SuSiE-IBOSS in settings with correlated covariates.

\begin{figure}[t]
\centering	

\begin{minipage}{0.49\textwidth}
    \centering
    \text{\footnotesize Fixed 12 signal loci }
\end{minipage}
\hfill
\begin{minipage}{0.49\textwidth}
    \centering
    \text{\footnotesize Random 50 signal loci }
\end{minipage}
\vspace{0.0em}
\begin{subfigure}{.50\textwidth}
		\centering
		\includegraphics[width=.9\linewidth, height=0.3\textheight, keepaspectratio]{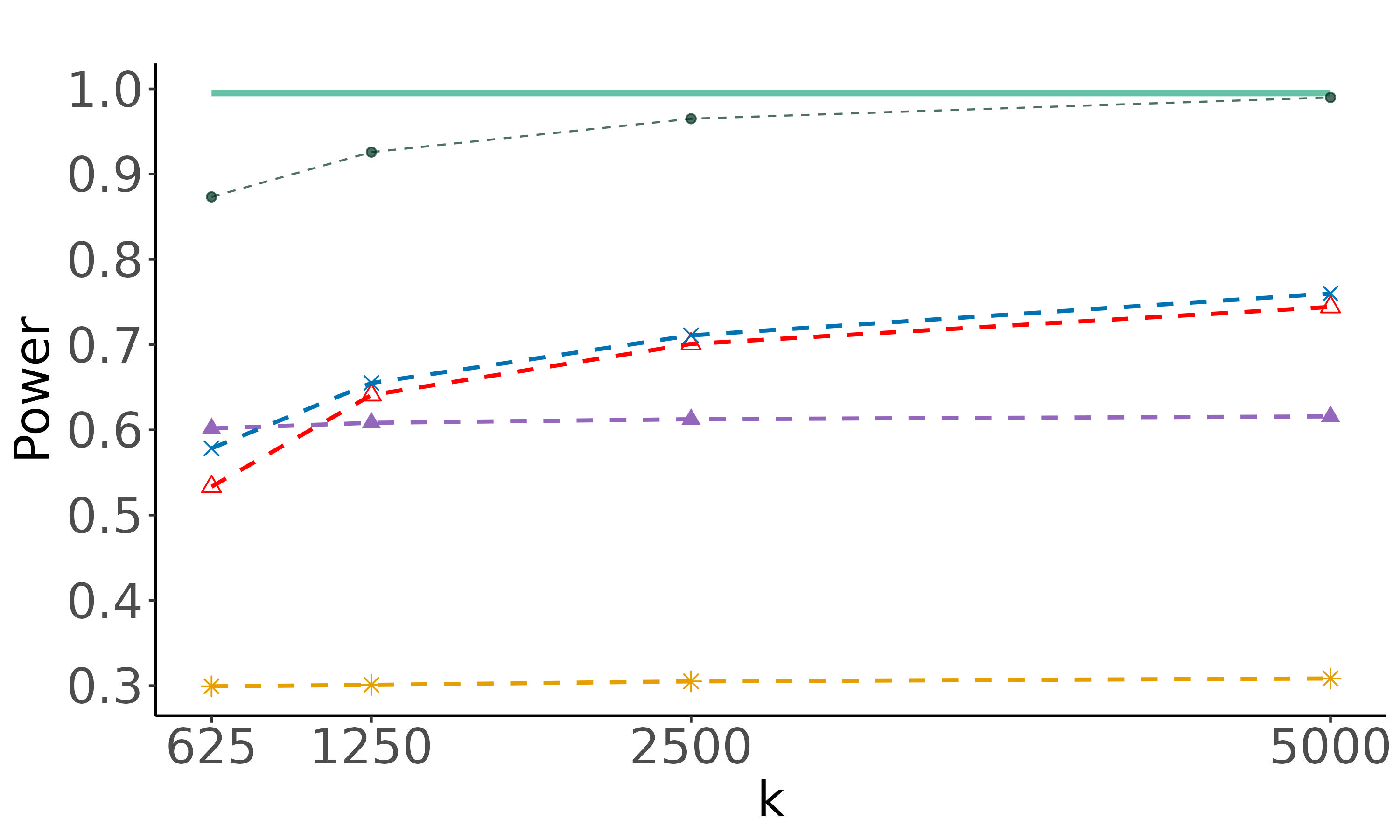}
	\end{subfigure}
    \hspace{-0.5em}
    \begin{subfigure}{.50\textwidth}
		\centering
		\includegraphics[width=.9\linewidth, height=0.3\textheight, keepaspectratio]{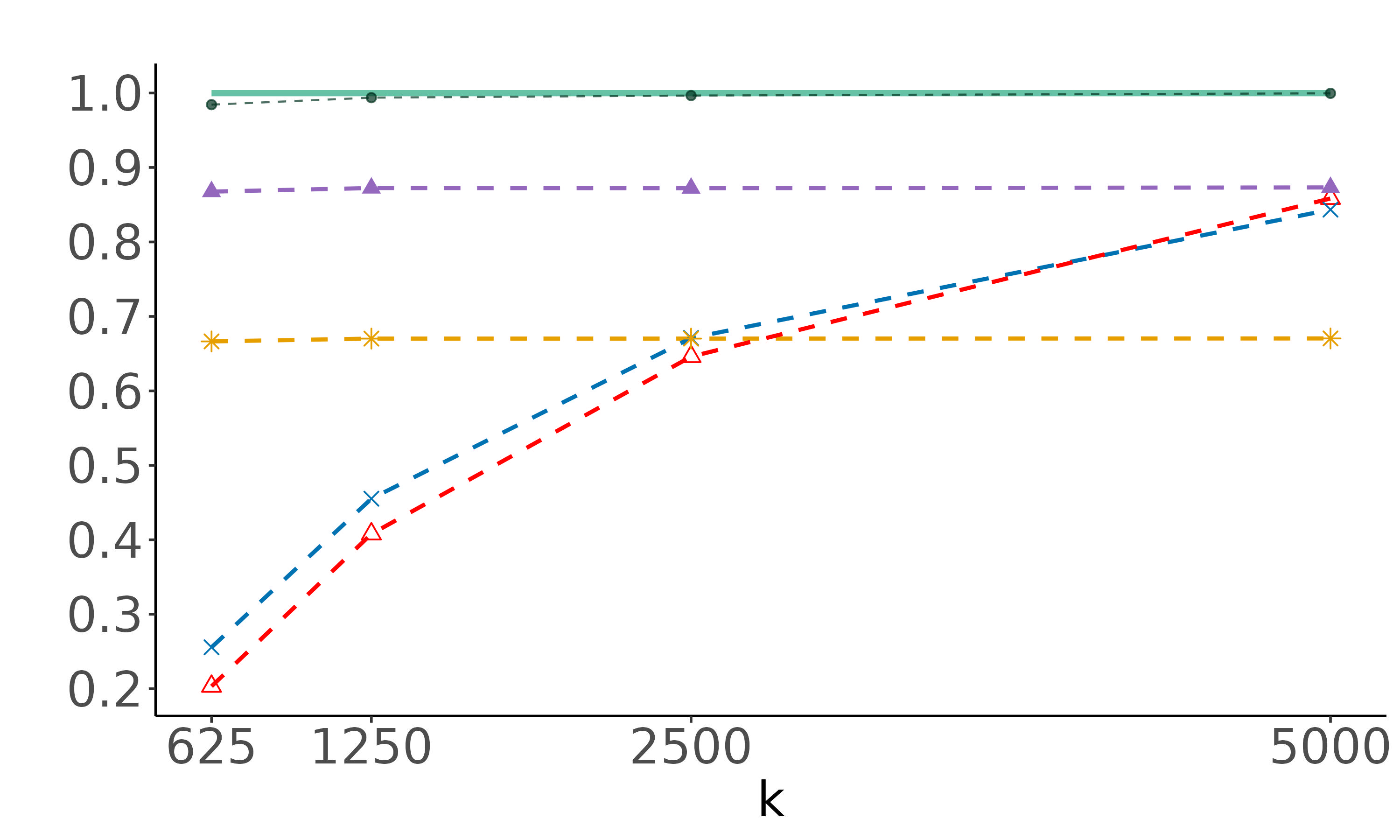}
	\end{subfigure}
{ \it \text{(a) Average variable-selection power }  }	

 \vspace{1.0em}

\centering
\begin{minipage}{0.49\textwidth}
    \centering
    \text{\footnotesize Fixed 12 signal loci }
\end{minipage}
\hfill
\begin{minipage}{0.49\textwidth}
    \centering
    \text{\footnotesize Random 50 signal loci }
\end{minipage}
\vspace{0.0em}
\begin{subfigure}{.50\textwidth}
		\centering
        \includegraphics[width=0.9\linewidth, height=0.3\textheight, keepaspectratio]{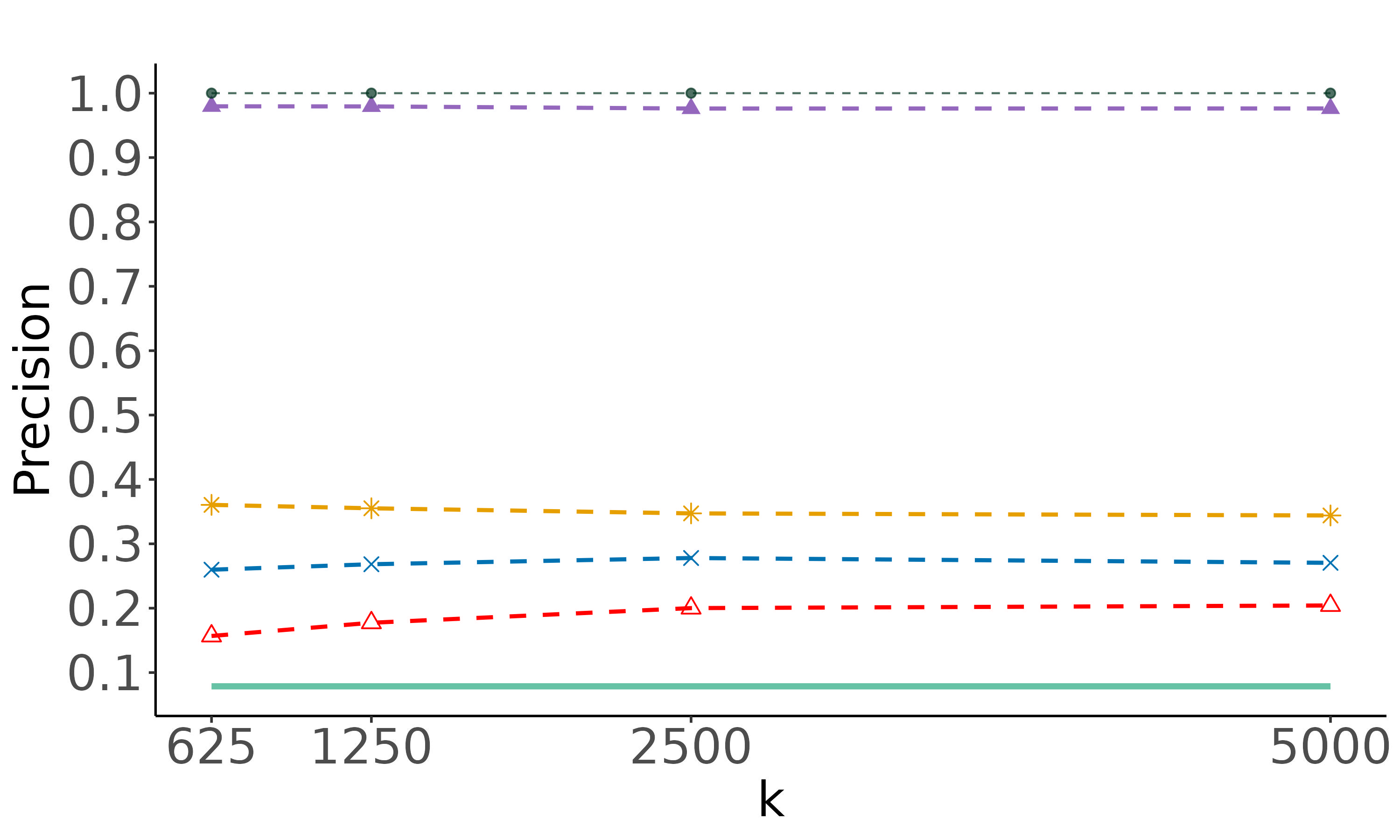}
	\end{subfigure}%
      \hspace{-0.5em}
   \begin{subfigure}{.50\textwidth}
		\centering
		\includegraphics[width=.9\linewidth, height=0.3\textheight, keepaspectratio]{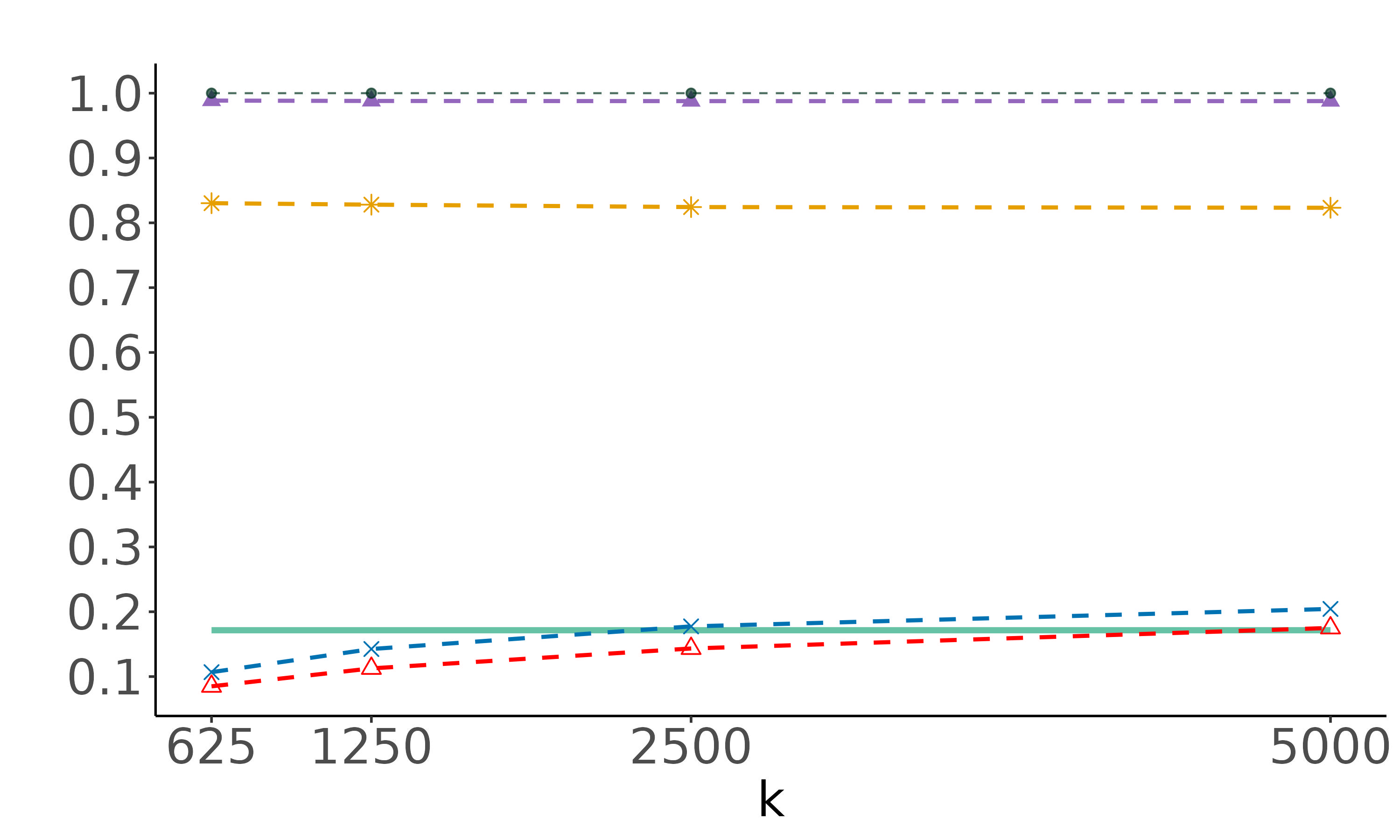}
	\end{subfigure}
      \hspace{-0.5em}
	
  { \it \text{(b) Average variable-selection precision }}
\vspace{1.0em} 

    \includegraphics[width=0.95\textwidth]{Plots/by_k_legend_only.png} 


\vspace{0.0em}

\caption{\it Benchmarking final variable selection after LASSO model fitting, based on $100$ simulation runs for the linear regression model with $\rho = 0.95$, $SNR = 0.5$, $n = 10^5$, $n_{\mathrm{test}}=30{,}000$, $p = 5000$, and $c = 0.5$.}
	\label{figure: Final variable selection} 
\end{figure}

Figure~\ref{figure:instance-comparisons} shows that oracle variable selection followed by IBOSS and LASSO achieves predictive performance close to full-data LASSO even with small subsample sizes of $k=1250$ and $2500$ in both settings. This illustrates that accurate recovery of the true signals in Stage I (better prescreening) can improve the effectiveness of informative subsampling and yield high predictive performance even for very small subsample sizes. 

Additional analyses in Supplementary Section~S.2 examine the contribution of informative subsampling in Stage~II to predictive performance. For the same Stage~I prescreening, IBOSS is compared with uniform random sampling in Supplementary Figure~S.2. Following each of the TH, CT, and SuSiE prescreening procedures, IBOSS achieves higher average test $r^2$ than uniform subsampling, with the differences being most apparent at smaller subsample sizes.

Although PTS is not specifically designed for signal recovery, we additionally evaluate final variable selection performance as a secondary criterion after LASSO model fitting on $\mathcal{D}_{I^\prime J^\prime}$ for PTS instances, $\mathcal{D}_{I^\prime J}$ for uniform subsampling, and $\mathcal{D}_{I J}$ for full-data training. Figure~\ref{figure: Final variable selection} shows that in the `Fixed 12 signal loci' setting, SuSiE-IBOSS has relatively stable power across subsample sizes, although TH-IBOSS and uniform subsampling achieve higher power at larger subsample sizes. In the `Random signal loci' setting, SuSiE-IBOSS achieves higher power than the other non-oracle subsampling methods at smaller subsample sizes. SuSiE-IBOSS also achieves substantially higher precision than the other non-oracle methods, including full-data LASSO. The CT-based PTS instance performs poorly overall in these settings, consistent with its poorer Stage I prescreening performance in the presence of highly correlated covariates. 

Additional simulations with moderately correlated covariates ($\rho=0.5$) are presented in Supplementary Section~S.4. The relative performance of the PTS variants depends on the signal configuration (see Figure~S.4). Overall, SuSiE-IBOSS is well suited for scalable model training in sparse settings with highly correlated covariates, achieving predictive performance comparable to the full-data model while substantially reducing runtime.

\section{An application to polygenic risk scoring} \label{Section: HAPNEST Data Example}

This section illustrates the performance of PTS for polygenic risk score (PRS) modeling using biobank-scale simulated HAPNEST genotype data. HAPNEST provides biobank-scale simulated individual-level genotype data comprising $6.8$ million common variants across the 22 autosomal chromosomes for $1,008,000$ individuals from six different ancestries \citep{HAPNEST_Paper}, together with downloadable PLINK-format files. This dataset has been shown to preserve key statistical properties of real genotype data, including LD structure. We consider training a PRS model for a continuous phenotype. As PRSs are known to be ancestry-specific, in this paper, we restrict the analysis to the European-ancestry cohort, comprising \(168,000\) individuals. We use SNPs on chromosome 22 with minor allele frequency (MAF) $\geq 1\%$, resulting in $ p = 107,598$ variants. For this analysis, the sparsity is set to $0.1\%$, corresponding to 107 true signals at randomly selected SNP positions with scaled Rademacher effects taking values \(-0.5\) and \(0.5\) with equal probability, similar to the `Random signal loci' setting in Section~\ref{Section: Numerical Results}. The phenotype is generated according to the linear regression model in \eqref{eq:linear_model} with a signal-to-noise ratio of $0.5$, corresponding to a heritability of $33.3\%$. 

Model fitting with this scale of data is computationally demanding and poses substantial memory challenges. For example, fitting the LASSO on the full data using the popular \texttt{cv.glmnet} implementation in \texttt{R} exceeds practical memory limits. Using \texttt{snpnet}, a specialized and computationally efficient implementation of LASSO for genotype data modeling, also requires considerable computational time. Therefore, we apply PTS with $k=5000$ to enable scalable PRS model training with the standard LASSO implementation in \texttt{cv.glmnet}. 
As an additional benchmark, we consider model training using a uniform random subsample of the same size ($k=5000$). For model training and evaluation, we use a $60\%-20\%-20\%$ split of the full dataset into training, validation, and test sets. This leads to a training data size of $n = 100800$, and test and validation data sizes of $n_\mathrm{test} = n_\mathrm{val} = 33600$. Both prescreening and subsampling are performed exclusively on the training set. This split is primarily used to accommodate the \texttt{snpnet} implementation in the benchmark comparison, as \texttt{snpnet} uses a separate validation set for tuning parameter selection rather than standard 10-fold cross-validation. The analysis results are based on 100 simulation runs. Across simulation replicates, the HAPNEST genotype data are held fixed, while the causal SNP locations, coefficient signs, phenotype errors, and training-validation-test split are regenerated. The exception is \texttt{snpnet}, which is evaluated in a single run because of its considerably higher computational cost.

\subsection{PTS implementation details}
In this application, the high dimensionality makes calculating marginal correlations for each variable in TH and subsequently performing clumping for CT computationally inefficient in \texttt{R}. We therefore use PLINK~\citep{chang2015second}, which provides computationally efficient implementations of these procedures for genotype data, and call it from the main \texttt{R} code to perform TH and CT in Stage I of PTS. The TH implementation is slightly modified from the one used in Section~\ref{Section: Numerical Results}, because the Bonferroni threshold retains approximately 5,000--8,000 variants in Stage I (across the 100 simulations), which would preclude the use of small subsamples under the standard IBOSS allocation, as this requires $k \ge 2q $. Therefore, we retain only the top 1,000 variants with the smallest marginal $p$-values. CT was performed in PLINK using a Bonferroni-corrected significance threshold of $0.05/p$ for index-variant selection and an LD $r^2$ threshold of $0.1$ for clumping within $\pm 250$~kb of each index variant. For Stage I SuSiE prescreening, the correlated blocks are defined as the LD regions of chromosome 22 using the predefined LD blocks available in the \texttt{R} package \texttt{BRIER}~\citep{berisa2015approximately}. The remaining implementation follows the procedure described in Section~\ref{Section: Numerical Results}, with one extension. Because of the more complex correlation structure among covariates, in addition to variants with PIP \(> 0.5\), the variant with the highest PIP from each SuSiE credible set (as a representative of variants in the credible set) is retained in Stage I prescreening. For the PTS instances, we set $c = 0.5$.

\subsection{Results}

Table~\ref{tab:HAPNEST_results} shows that, in this HAPNEST data setting, SuSiE-IBOSS and CT-IBOSS provide the strongest overall performance. SuSiE-IBOSS and CT-IBOSS achieve test $r^2$ values of $0.309$ and $0.295$ using approximately $88$ and $290$ prescreened SNPs, respectively. Both perform close to the oracle-IBOSS benchmark and the full-data \texttt{snpnet} fit (single run), which attain test $r^2$ values of $0.320$ and $0.325$, respectively. SuSiE-IBOSS and CT-IBOSS have total runtimes of 402 and 20 seconds, respectively, both lower than the  551 seconds required for uniform subsampling. The runtime of the full-data \texttt{snpnet} fit based on a single validation set is 4740 seconds, substantially higher than that of SuSiE-IBOSS. 
Additionally, SuSiE-IBOSS and CT-IBOSS have peak memory requirements of $9.1$ GB and $2.2$ GB, respectively, compared with $40$ GB for uniform subsampling and $16$ GB for the full-data \texttt{snpnet} fit. 
TH-IBOSS shows substantially lower predictive performance in this setting, consistent with less effective prescreening under the complex LD structure and high dimensionality. 
Uniform subsampling also yields lower predictive performance, with a test $r^2$ of $0.239$, while requiring more runtime and memory than SuSiE-IBOSS and CT-IBOSS.

\begin{table}[htbp]
\centering
\caption{Average predictive and computational performance for subsamples of size $k=5000$ over 100 simulation runs. PTS instances use TH, CT, SuSiE, or oracle prescreening in Stage I, followed by IBOSS subsampling and LASSO model training. Uniform subsampling uses no prescreening and LASSO is fitted using all SNPs.}

\begin{tabular}{lcccc|cc}
\toprule

Method & SNPs for fitting & Test $r^2$ & Runtime (s) & Peak memory (GB) &  Final power & Final precision \\
\midrule

Oracle-IBOSS & 107    & 0.320 & 4     &  1.6    &    0.99 & 1.00 \\
TH-IBOSS    & 1,000   & 0.201 & 245 &  1.3    & 0.20 & 0.21 \\
CT-IBOSS    & 290   & 0.295 & 20   &  2.2    & 0.53 & 0.32 \\
SuSiE-IBOSS & 88     & 0.309 & 402 &   9.1   & 0.54 & 0.66 \\
\midrule
Uniform     & 107,598 & 0.239 & 551 &  40 &    0.39 & 0.07 \\

\bottomrule
\end{tabular}
\label{tab:HAPNEST_results}
\end{table}

Overall, the PRS analysis demonstrates that SuSiE-IBOSS and CT-IBOSS enable scalable model fitting for genotype data. Compared with uniform subsampling and the full-data \texttt{snpnet} fit, they require substantially less memory and shorter computation times. Additionally, their predictive performance is comparable to that of the full-data \texttt{snpnet} fit. Notably, CT-IBOSS performs substantially better in the PRS analysis than in the simulation settings of Section~\ref{Section: Numerical Results}. The PRS setting differs in several aspects, including substantially higher dimensionality, a sparser signal structure ($s =0.1\%$), and a more realistic LD structure. The SuSiE-IBOSS and CT-IBOSS implementations of the PTS algorithm are particularly effective for sparse, high-dimensional data with correlated covariates, making them well suited for large-scale PRS modeling.

\section{Discussion}

In this work, we integrate prescreening with informative subsampling for scalable model training in high-dimensio\-nal and large-sample data. Across the correlated Gaussian simulation settings and the HAPNEST genotype application, PTS with multivariable SuSiE prescreening substantially reduced runtime while maintaining predictive performance close to full-data training. In the large-scale HAPNEST setting for polygenic risk score modeling, PTS additionally provided considerable reductions in memory requirements.

The effectiveness of PTS depends on the quality of the prescreening step, as informative subsampling in Stage II is performed within the covariate space identified in Stage I. If relevant variables are omitted, informative directions are no longer available to guide subsampling or contribute to the final prediction model, while retaining non-informative or redundant variables can distribute some of the available subsampling effort across irrelevant directions. These two mechanisms are illustrated in Proposition~\ref{Propsition PTS}. The importance of Stage~I is further reflected in the oracle simulations, where using the true active set in Stage I leads to high predictive performance even for very small subsample sizes. Motivated by this, we incorporate multivariable prescreening based on individual-level data to account for the joint signal structure among correlated covariates. 

In our simulations, multivariable prescreening using SuSiE achieved substantially higher precision than margi\-nal screening using thresholding~(TH) while retaining relatively high power, and SuSiE-IBOSS showed improved predictive performance compared with TH-IBOSS, especially for small subsample sizes. The supplementary comparison with uniform subsampling further shows that informative IBOSS selection provides an additional gain beyond prescreening, especially when the subsample size is small. Considering the results across Sections~\ref{Section: Numerical Results} and~\ref{Section: HAPNEST Data Example}, SuSiE-IBOSS appears particularly well suited for sparse, high-dimensional data with correlated covariates. The HAPNEST results further show that CT-IBOSS can achieve strong predictive performance despite lower variable selection precision, while requiring less runtime and memory than SuSiE-IBOSS.

Recent approaches such as SIS-IBOSS \citep{wang2024efficient}, CLASS \citep{singh2023subdata}, FAME \citep{wei2024efficient} and the approach of \citet{chasiotis2026efficient} also combine feature selection with observation reduction for large-scale data. FAME combines marginal feature screening with information-based observation reduction, while \citet{chasiotis2026efficient} combine LASSO-based variable selection with leverage-score subsampling. PTS complements these approaches by allowing the prescreening step to explicitly account for covariate correlation, including via multivariable fine-mapping with SuSiE, before informative subsampling.

A potential limitation is that the effectiveness of the present PTS implementation depends on the dimension of the prescreened covariate space. In particular, the standard IBOSS allocation requires the number of retained covariates~\(q\) to be sufficiently small relative to the target subsample size~\(k\) (\(k\geq 2q\)); consequently, less sparse settings may require larger subsamples or more aggressive prescreening. Future work will focus on developing subsampling strategies that are more robust to prescreening errors and model misspecification. As a further direction, we are exploring whether samples can instead be identified using a small number of derived variables, such as principal components (PCs). The HAPNEST analysis provides a biobank-scale evaluation under a realistic LD structure and extends the Gaussian simulation study to a substantially more complex covariate structure; however, it remains based on simulated phenotypes and is restricted to chromosome 22 and individuals of European ancestry. Future work will evaluate PTS with observed phenotypes and genome-wide data from large cohort studies such as UK Biobank and the German National Cohort (NAKO). Although PTS is formulated for general exponential-family regression models, the present numerical evaluation focuses on linear models with IBOSS for informative subsampling. Evaluating PTS for other outcome types and alternative subsampling methods provides another direction for future research.

\section*{Data availability and computational environment}

The HAPNEST dataset used in this study is publicly available from the EMBL--EBI BioStudies repository under accession \texttt{S-BSST936}:
\url{https://www.ebi.ac.uk/biostudies/studies/S-BSST936}. All computations were performed on an HPC Linux server running Ubuntu~24.04.4~LTS, equipped with two AMD EPYC~7713 processors and 1~TB of RAM. The R code accompanying this paper is available at \url{https://github.com/Subhadrad/PTS}. 

\section*{Acknowledgments}

This work was supported in part by the Deutsche Forschungsgemeinschaft (DFG, German Research Foundation) -- Project numbers 587908135 and 427806116 (Research Training Group 2624, ``Biostatistical Methods for High-Dimensional Data in Toxicology'').

\section*{Declaration of generative AI use}

OpenAI's ChatGPT (GPT-5.6 and GPT-6) was used interactively to assist with language editing, programming tasks (code suggestions and debugging), and the iterative development, refinement, and checking of the mathematical formulation and proof of Proposition~\ref{Propsition PTS}. All AI-assisted suggestions were critically evaluated, substantially refined, scientifically contextualized, and independently verified by the authors, who take full responsibility for the scientific content of the manuscript.


\bibliography{References} 

\clearpage 

\setcounter{section}{0}
\renewcommand{\thesection}{S.\arabic{section}}

\setcounter{figure}{0}
\renewcommand{\thefigure}{S.\arabic{figure}}

\setcounter{table}{0}
\renewcommand{\thetable}{S.\arabic{table}}

\begin{center}
    {\LARGE\bfseries Supplementary Material\par}
\end{center}
\vspace{1em} 

\section{Effect of the prescreening sample fraction~$c$} \label{Supplementary Section 1}

The prescreening sample fraction~$c$ affects both predictive performance and computational efficiency. For very large datasets, such as biobank-scale applications with \(n\) on the order of \(10^5\) or larger, using only a fraction of the observations in Stage I may still provide sufficient information for effective prescreening, while substantially reducing computational cost. This can be particularly beneficial for computationally intensive fine-mapping methods such as SuSiE.  
We consider the same linear regression simulation setting in Section~3 of the main paper with $n=100{,}000$, $p=5000$ ($B=5$ and $m=1000$), $\rho=0.95$, $\mathrm{SNR}=0.5$, $\beta_0=1$, under both the `Fixed 12 signal loci' and the `Random signal loci' setting ($s=1\%$). To investigate the sensitivity of Stage~I prescreening to the initial sample fraction, we vary $c\in\{1,0.75,0.5,0.25\}$, corresponding to $100\%$, $75\%$, $50\%$, and $25\%$ of the initial training observations, respectively.

\begin{figure}[t]
\centering	
\begin{minipage}{0.49\textwidth}
    \centering
    \text{\footnotesize Fixed 12 signal loci }
\end{minipage}
\hfill
\begin{minipage}{0.49\textwidth}
    \centering
    \text{\footnotesize Random 50 signal loci }
\end{minipage}
\vspace{0.0em}
\begin{subfigure}{.50\textwidth}
		\centering
		\includegraphics[width=.9\linewidth, height=0.3\textheight, keepaspectratio]{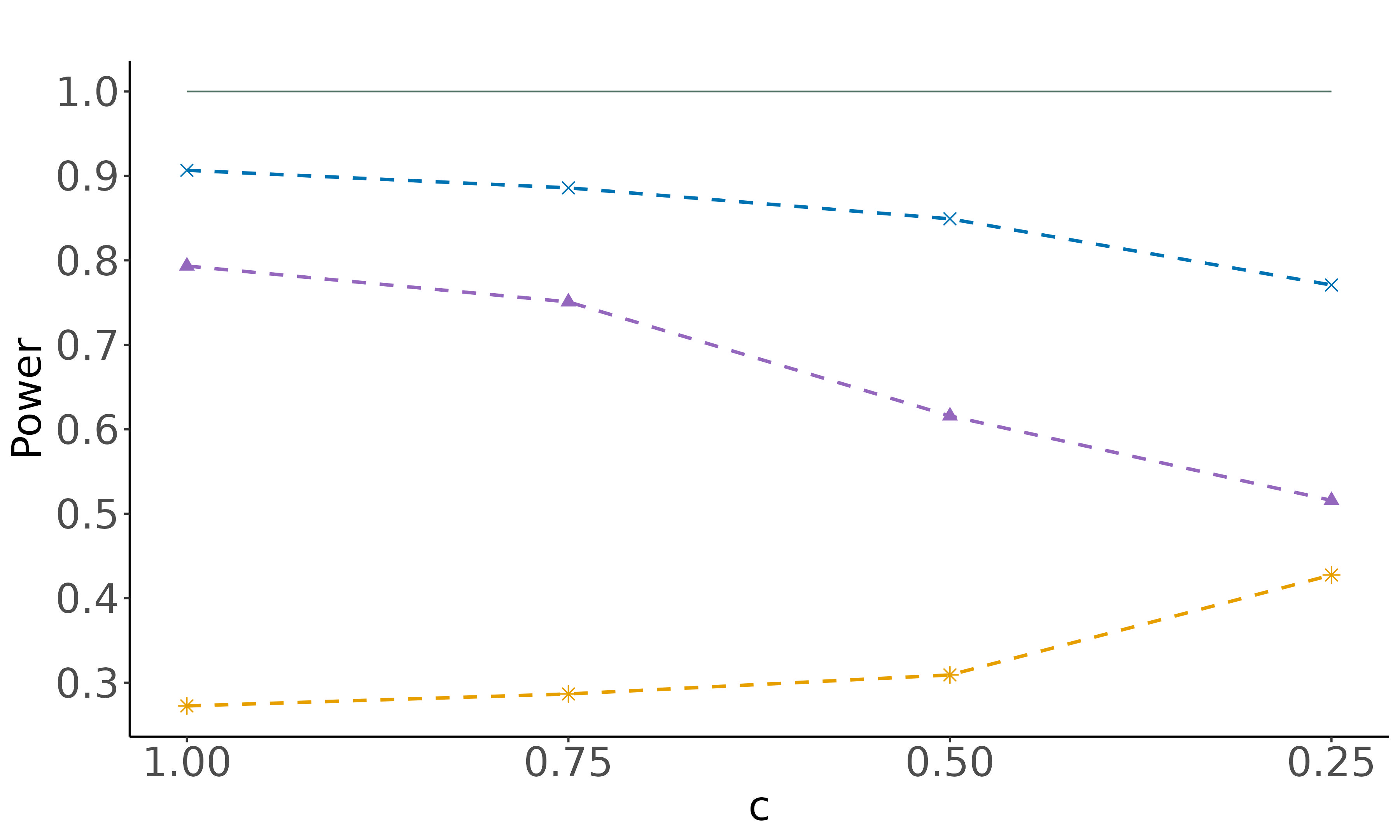}
	\end{subfigure}
    \hspace{-0.5em}
    \begin{subfigure}{.50\textwidth}
		\centering
		\includegraphics[width=.9\linewidth, height=0.3\textheight, keepaspectratio]{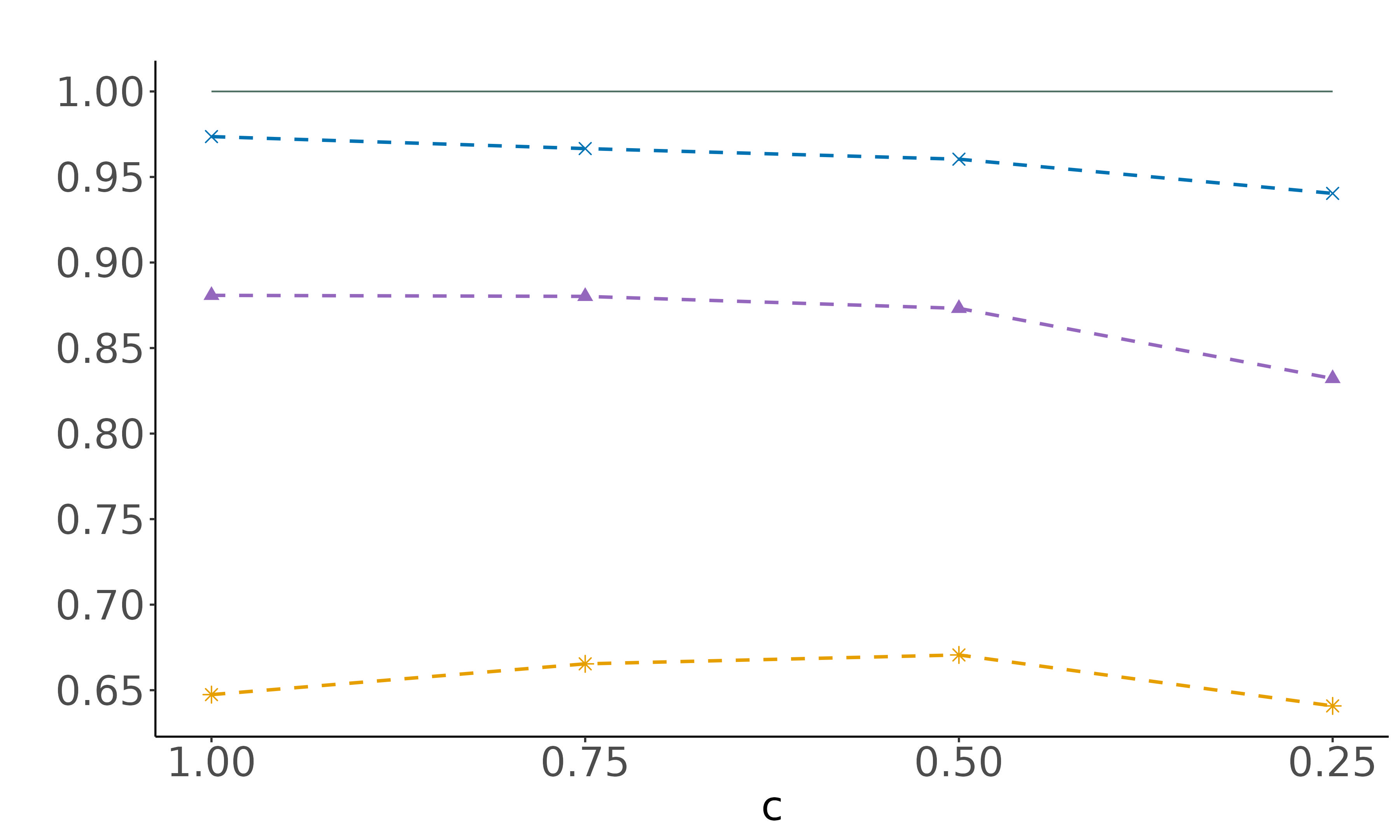}
	\end{subfigure}
{ \it \text{(a) Average Stage I variable-selection power }  }	
\vspace{1.0em}

\centering
\begin{minipage}{0.49\textwidth}
    \centering
    \text{\footnotesize Fixed 12 signal loci }
\end{minipage}
\hfill
\begin{minipage}{0.49\textwidth}
    \centering
    \text{\footnotesize Random 50 signal loci }
\end{minipage}
\vspace{0.0em}
\begin{subfigure}{.50\textwidth}
		\centering
\includegraphics[width=0.9\linewidth, height=0.3\textheight, keepaspectratio]{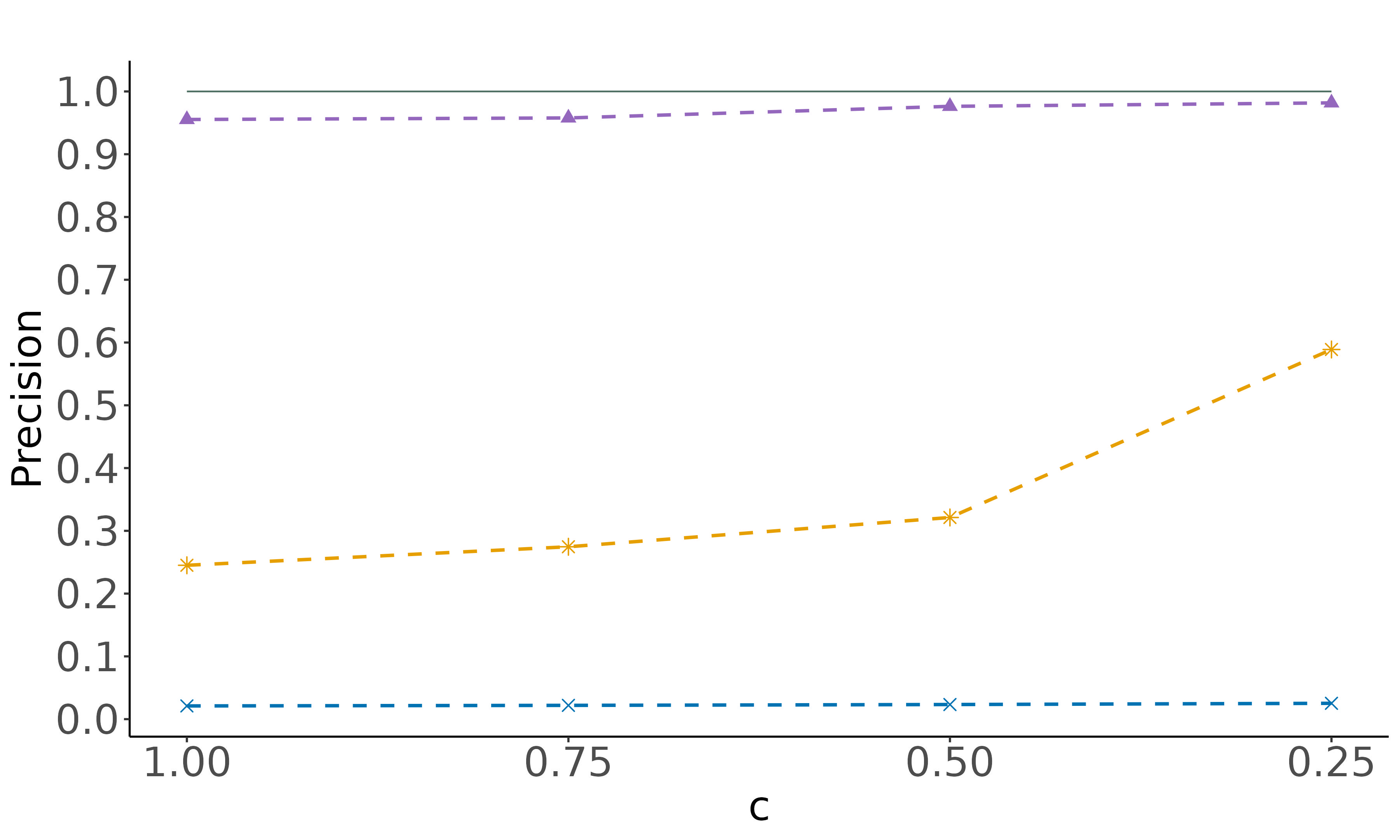}
	\end{subfigure}%
      \hspace{-0.5em}
   \begin{subfigure}{.50\textwidth}
		\centering
		\includegraphics[width=.9\linewidth, height=0.3\textheight, keepaspectratio]{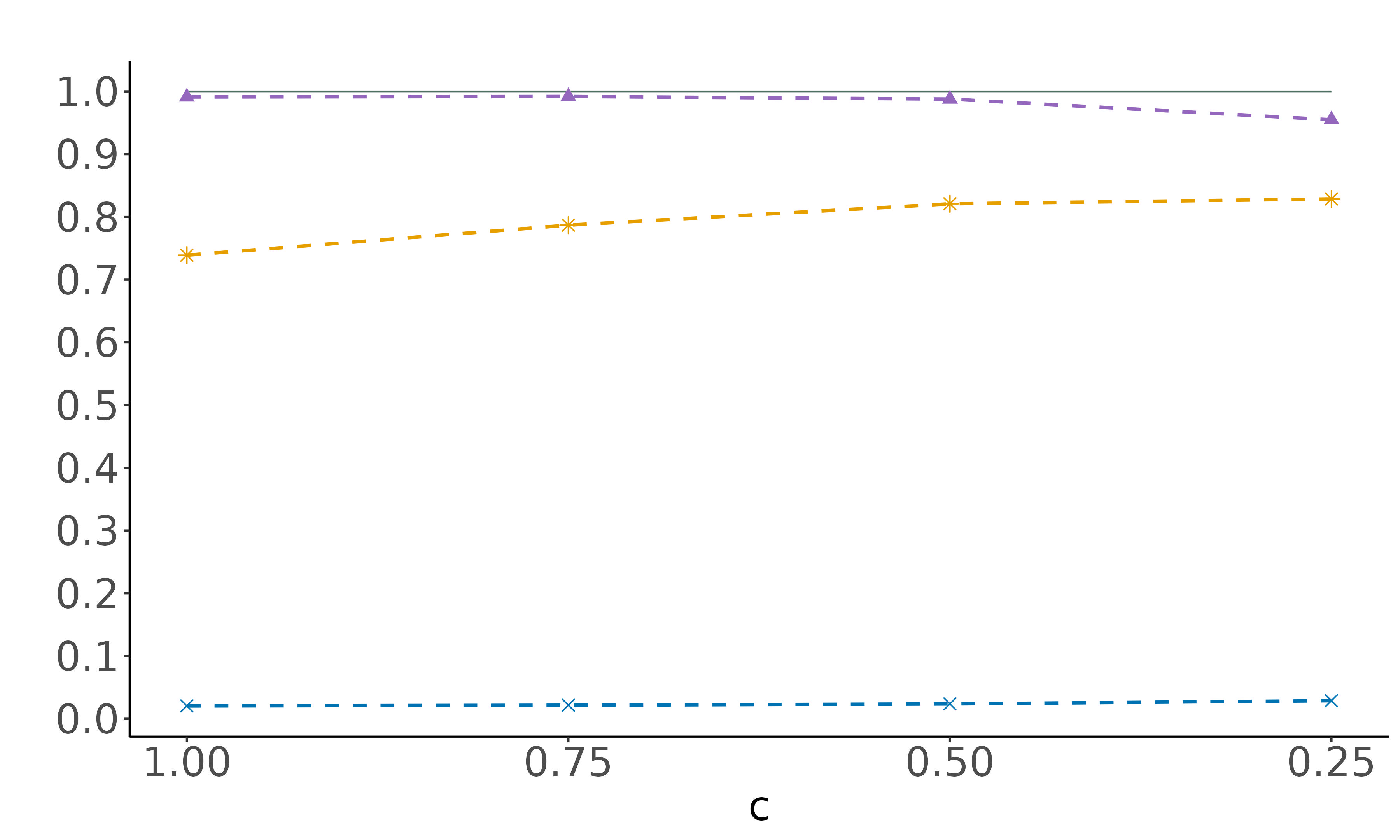}
	\end{subfigure}
      \hspace{-0.5em}
	
  { \it \text{(b) Average Stage I variable-selection precision }}
\vspace{1.0em} 

    \includegraphics[width=0.95\textwidth]{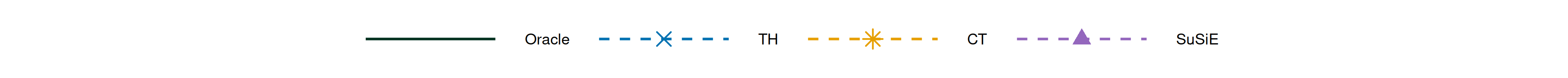}

\vspace{0.0em}

\caption{\it Benchmarking initial variable selection, after running the prescreening algorithms in Stage I of PTS, based on $100$ simulation runs for the linear regression model with $\rho = 0.95$, $SNR = 0.5$, $n=100{,}000$, $p = 5000$, and $c\in\{0.25,0.5,0.75,1\}$.}
	\label{Figure: Initial variable selection} 
\end{figure}

Figure~\ref{Figure: Initial variable selection} shows that prescreening performance is relatively stable across $c=1,\,0.75,$ and $0.5$ in the `Random 50 signal loci' setting, whereas SuSiE power decreases more noticeably with decreasing $c$ in the `Fixed 12 signal loci' setting. We therefore use $c=0.5$ in our simulation studies as a compromise between retaining effective prescreening performance and reducing the computational cost of Stage~I.  This reduction is particularly advantageous for computationally intensive SuSiE prescreening in large-scale settings.

Figure~\ref{Figure: Initial variable selection} also shows that SuSiE achieves substantially higher precision (close to $1.0$) than TH for all prescreening sample fractions, although its power is lower, particularly in the `Fixed 12 signal loci' setting.  Unfortunately, clumping performs poorly in these settings, particularly for `Fixed 12 signal loci', where signals are deliberately placed adjacent or very close to one another. Since clumping removes nearby correlated variants, it fails to retain these adjacent signals, leading to lower power. Overall, SuSiE provides a favorable power-precision
tradeoff in these highly correlated settings. This is finally reflected in Figure 2 of the main paper, where we see that SuSiE-IBOSS generally leads to higher prediction accuracy.

\section{Additional analyses of PTS performance}
This section provides additional analyses that complement the numerical study in Section~3 of the main paper, focusing on uncertainty in the performance estimates, and on the role of Stage~II informative subsampling in the predictive performance and scalability of PTS. We consider the same linear regression simulation setting in Section~3 of the main paper with $100$ simulation runs, $c = 0.5$, $n=100{,}000$, $n_{\mathrm{test}}=30{,}000$, $p=5000$ ($B=5$ and $m=1000$), $\rho=0.95$, $\mathrm{SNR}=0.5$, $\beta_0=1$, under both the `Fixed 12 signal loci' setting and the `Random signal loci' setting ($s=1\%$). 

\subsection{Uncertainty in PTS performance estimates}

Figure~2 of the main paper reports the average $r_{test}^2$ over $100$ simulation replicates. Table~\ref{tab:r2_ci_combined} additionally reports $95\%$ confidence intervals for the expected  $r_{test}^2$ under the corresponding simulation setting, estimated from these replicates. The 95\% confidence intervals are narrow, indicating low Monte Carlo uncertainty in the estimated mean test $r^2$ values.

\begin{table}[ht]
\centering
\small
\caption{Average test $r^2$ with 95\% confidence intervals.}
\label{tab:r2_ci_combined}
\begin{tabular}{lcccc}
\toprule
& \multicolumn{4}{c}{$k$} \\
& 625 & 1250 & 2500 & 5000 \\
\midrule
\multicolumn{5}{c}{\textit{Fixed 12 signal loci}} \\
\midrule
Full Sample & 0.333 (0.332, 0.334) & 0.333 (0.332, 0.334) & 0.333 (0.332, 0.334) & 0.333 (0.332, 0.334) \\
Oracle IBOSS Subsample & 0.327 (0.326, 0.328) & 0.330 (0.329, 0.331) & 0.332 (0.331, 0.333) & 0.333 (0.332, 0.334) \\
Uniform Subsample & 0.289 (0.287, 0.292) & 0.312 (0.311, 0.314) & 0.323 (0.322, 0.324) & 0.328 (0.327, 0.328) \\
TH IBOSS Subsample & 0.309 (0.307, 0.311) & 0.321 (0.320, 0.322) & 0.326 (0.325, 0.327) & 0.329 (0.328, 0.330) \\
CT IBOSS Subsample & 0.192 (0.191, 0.193) & 0.195 (0.194, 0.196) & 0.197 (0.196, 0.197) & 0.197 (0.196, 0.198) \\
SuSiE IBOSS Subsample & 0.327 (0.326, 0.328) & 0.328 (0.327, 0.329) & 0.329 (0.328, 0.330) & 0.329 (0.328, 0.330) \\
\midrule
\multicolumn{5}{c}{\textit{Random 50 signal loci}} \\
\midrule
Full Sample & 0.332 (0.331, 0.333) & 0.332 (0.331, 0.333) & 0.332 (0.331, 0.333) & 0.332 (0.331, 0.333) \\
Oracle IBOSS Subsample & 0.293 (0.291, 0.295) & 0.313 (0.312, 0.314) & 0.323 (0.322, 0.324) & 0.328 (0.327, 0.329) \\
Uniform Subsample & 0.136 (0.132, 0.140) & 0.211 (0.208, 0.213) & 0.265 (0.263, 0.267) & 0.298 (0.297, 0.300) \\
TH IBOSS Subsample & 0.184 (0.180, 0.187) & 0.238 (0.236, 0.241) & 0.279 (0.277, 0.280) & 0.304 (0.303, 0.305) \\
CT IBOSS Subsample & 0.242 (0.239, 0.245) & 0.258 (0.256, 0.261) & 0.266 (0.264, 0.269) & 0.271 (0.268, 0.273) \\
SuSiE IBOSS Subsample & 0.282 (0.279, 0.284) & 0.299 (0.297, 0.301) & 0.308 (0.306, 0.310) & 0.313 (0.311, 0.314) \\
\bottomrule
\end{tabular}
\end{table}

\subsection{Informative sampling further enhances predictive performance}
In this section, we investigate the contribution of Stage~II informative subsampling to the predictive performance and scalability of PTS. To isolate the effect of Stage~II, we keep the Stage~I prescreening method fixed and consider TH, CT, and SuSiE separately. For each prescreening method, we compare the informative IBOSS subsample with a uniform random subsample of the same size, and fit the final model on each resulting subsample. 
\begin{figure}[H]
	\centering
	\begin{minipage}{0.49\textwidth}
    \centering
    \text{\footnotesize Fixed 12 signal loci }
\end{minipage}
\hfill
\begin{minipage}{0.49\textwidth}
    \centering
    \text{\footnotesize Random 50 signal loci }
\end{minipage}
\vspace{0.0em}
	\begin{subfigure}{.495\textwidth}
		\centering
		\includegraphics[width=.9\linewidth, height=0.3\textheight, keepaspectratio]{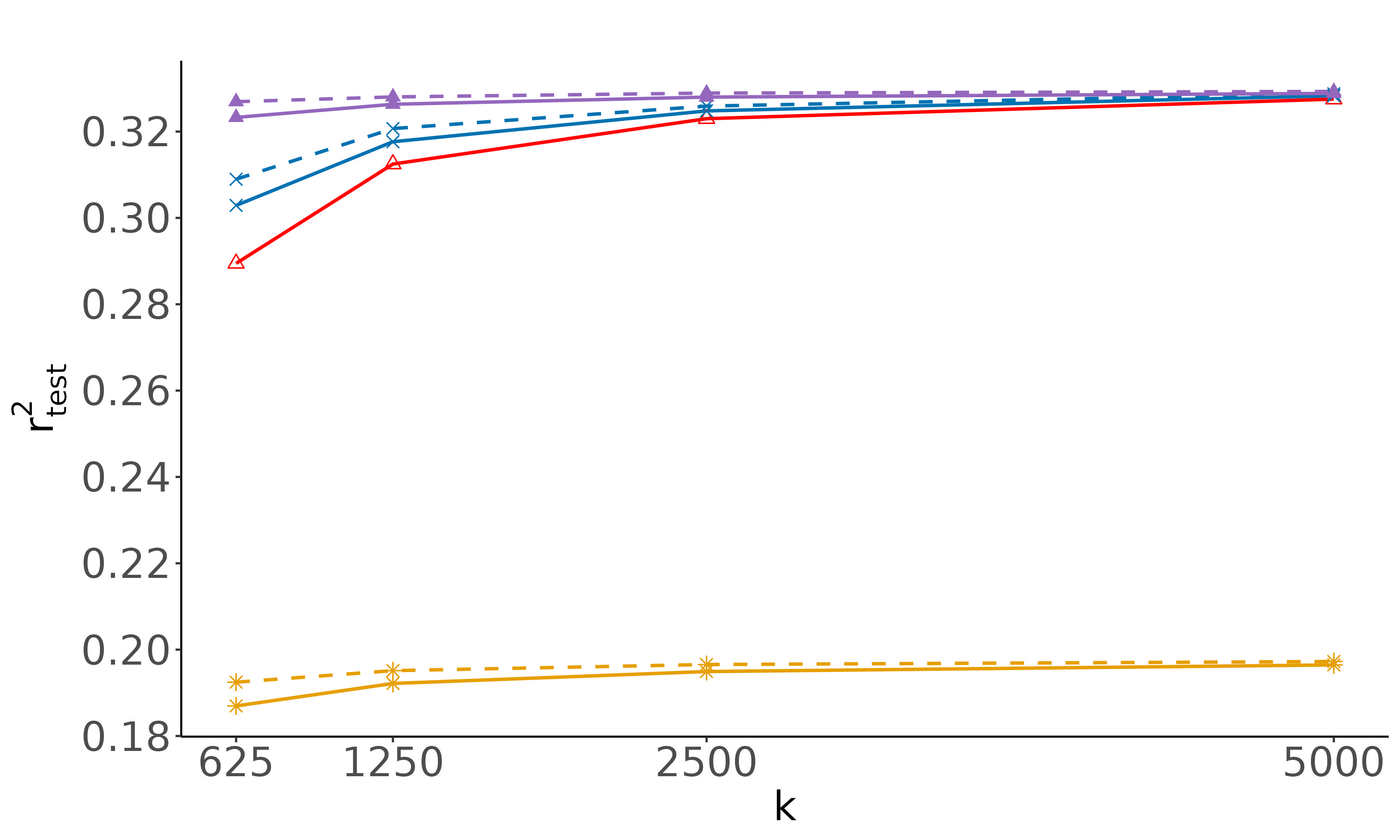}
	\end{subfigure}
    \begin{subfigure}{.495\textwidth}
		\centering
		\includegraphics[width=.9\linewidth, height=0.3\textheight, keepaspectratio]{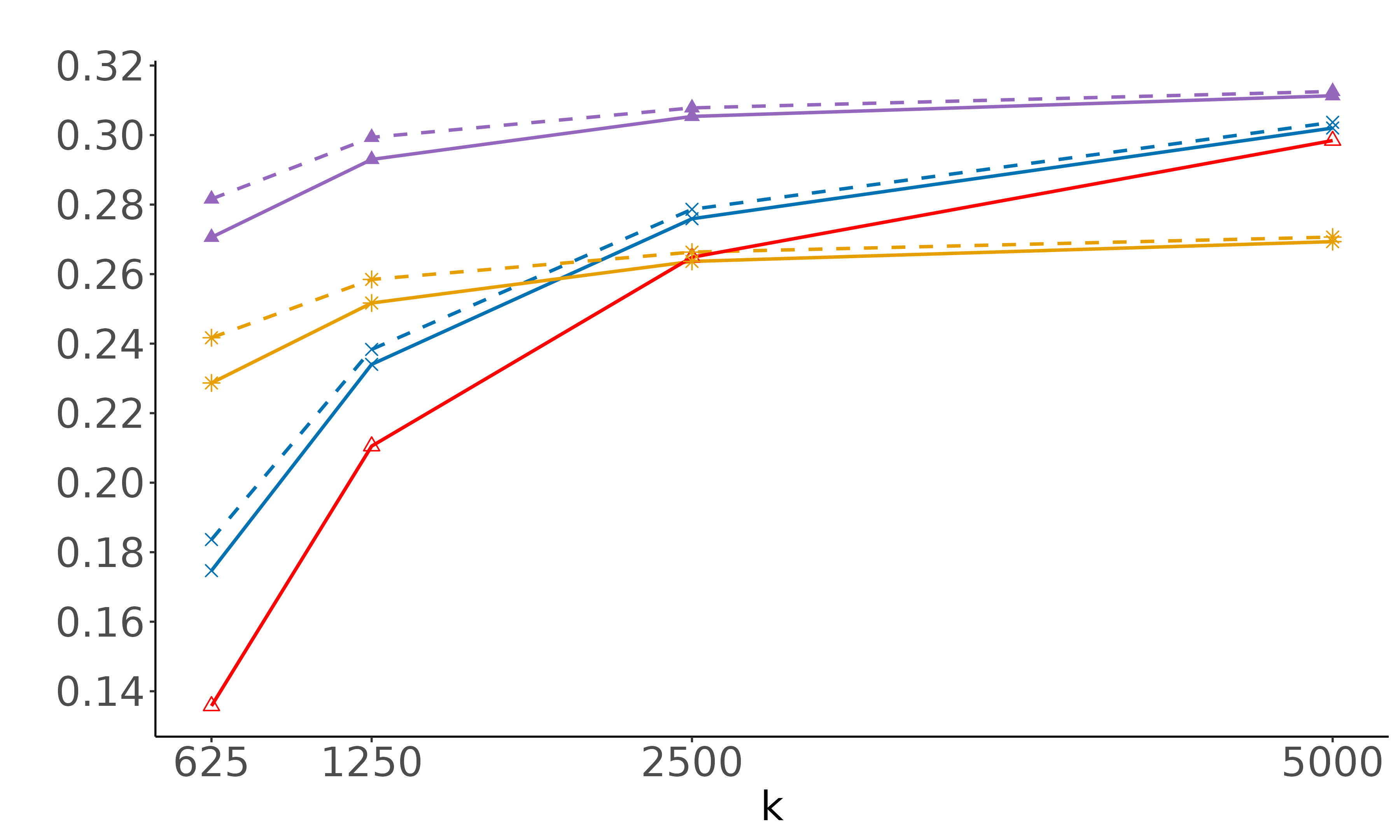}
	\end{subfigure}

   \includegraphics[width=0.95\textwidth]{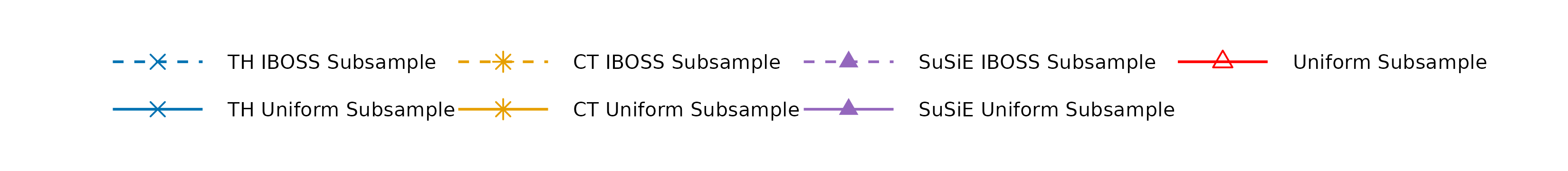} 

	\vspace{2em}
	\centering	\caption{ \it The simulated average test $r^2$ based on $100$ simulation runs, for linear regression model with $\rho = 0.95$, $SNR = 0.5$, $n=100{,}000$, $n_{\mathrm{test}}=30{,}000$, and $p = 5000$, with Stage~I prescreening performed on a random 50\% sample of the training observations ($c=0.5$). 
    }
	\label{Figure: Sensitivity Unif Vs IBOSS} 
\end{figure}
Figure~\ref{Figure: Sensitivity Unif Vs IBOSS} shows that, in both the “Fixed 12 signal loci” and “Random 50 signal loci” settings, informative IBOSS subsampling generally yields better predictive performance than uniform subsampling. The advantage is particularly pronounced for small subsample sizes, such as $k=625$ and $k=1250$, but becomes less evident at $k=5000$. Thus, informative subsampling in Stage II can be particularly valuable when subsample sizes are small.

\section{Sensitivity analysis}
TH-IBOSS can also be implemented using SIS prescreening, that is, by retaining the top \(q\) covariates ranked by absolute marginal correlation with the response. In the presence of highly correlated covariates, selecting the top \(q\) covariates does not guarantee retaining the truly active covariates, even those with large effect sizes. This can occur because inactive covariates near active covariates are strongly correlated with them and therefore also exhibit high marginal correlations with the response.

In this section, we conduct a sensitivity analysis to understand how many covariates need to be retained by the SIS implementation of TH-IBOSS in Stage I to achieve final predictive performance comparable to that of SuSiE-IBOSS. We consider the same linear regression simulation setting in Section~3 of the main paper with $n=100{,}000$, $n_{\mathrm{test}}=30{,}000$, $p=5000$ ($B=5$ and $m=1000$), $c=0.5$, $\rho=0.95$, $\mathrm{SNR}=0.5$, $\beta_0=1$, under both `Fixed 12 signal loci' and `Random 50 signal loci' ($s=1\%$), and conduct the analysis for $k=5000$.

\begin{figure}[H]
	\centering	
\begin{minipage}{0.49\textwidth}
    \centering
    \text{\footnotesize Fixed 12 signal loci }
\end{minipage}
\hfill
\begin{minipage}{0.49\textwidth}
    \centering
    \text{\footnotesize Random 50 signal loci }
\end{minipage}
\vspace{0.0em}

	\begin{subfigure}{.50\textwidth}
		\centering
		\includegraphics[width=\linewidth, height=0.3\textheight, keepaspectratio]{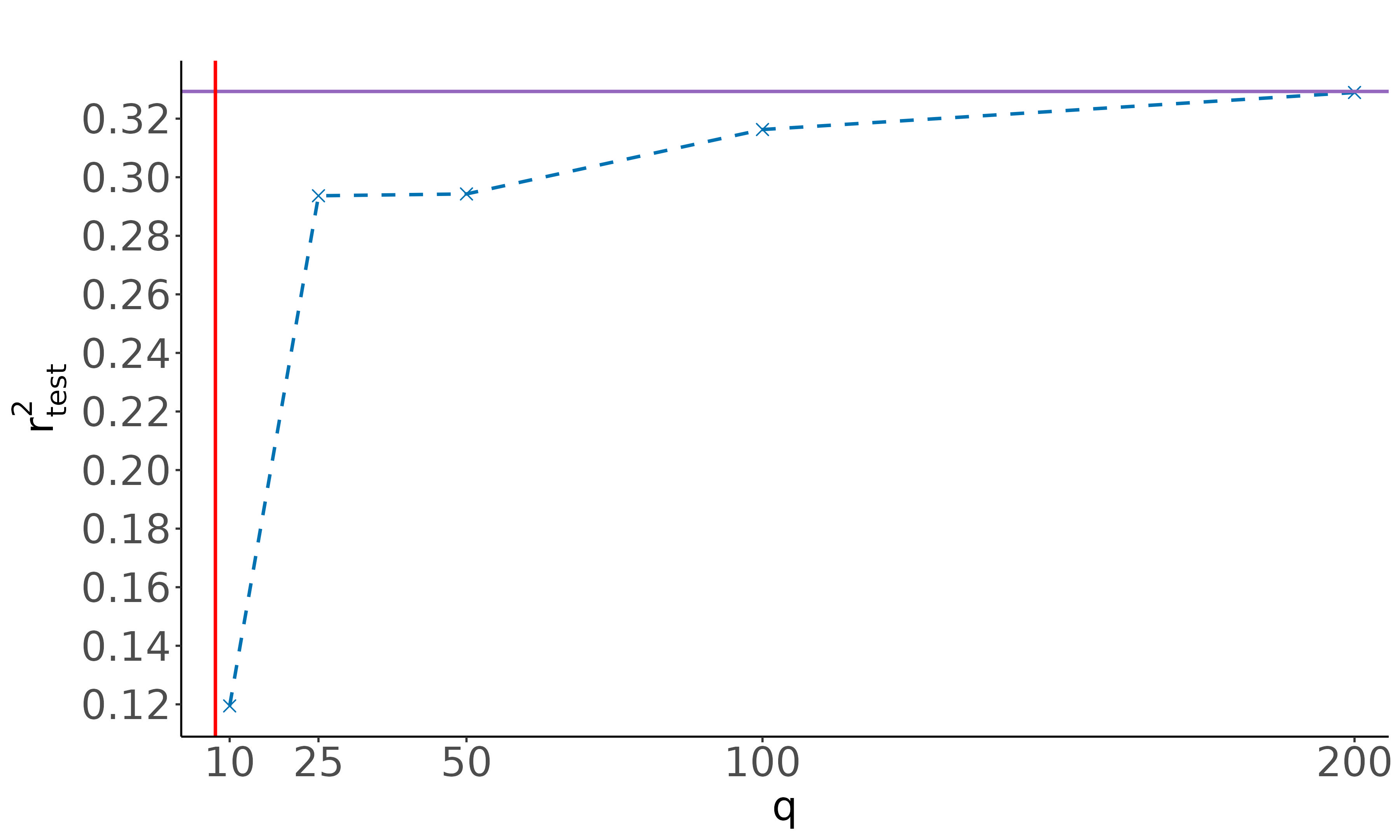}
	\end{subfigure}
      \hspace{-0.5em}
    \begin{subfigure}{.50\textwidth}
		\centering
		\includegraphics[width=\linewidth, height=0.3\textheight, keepaspectratio]{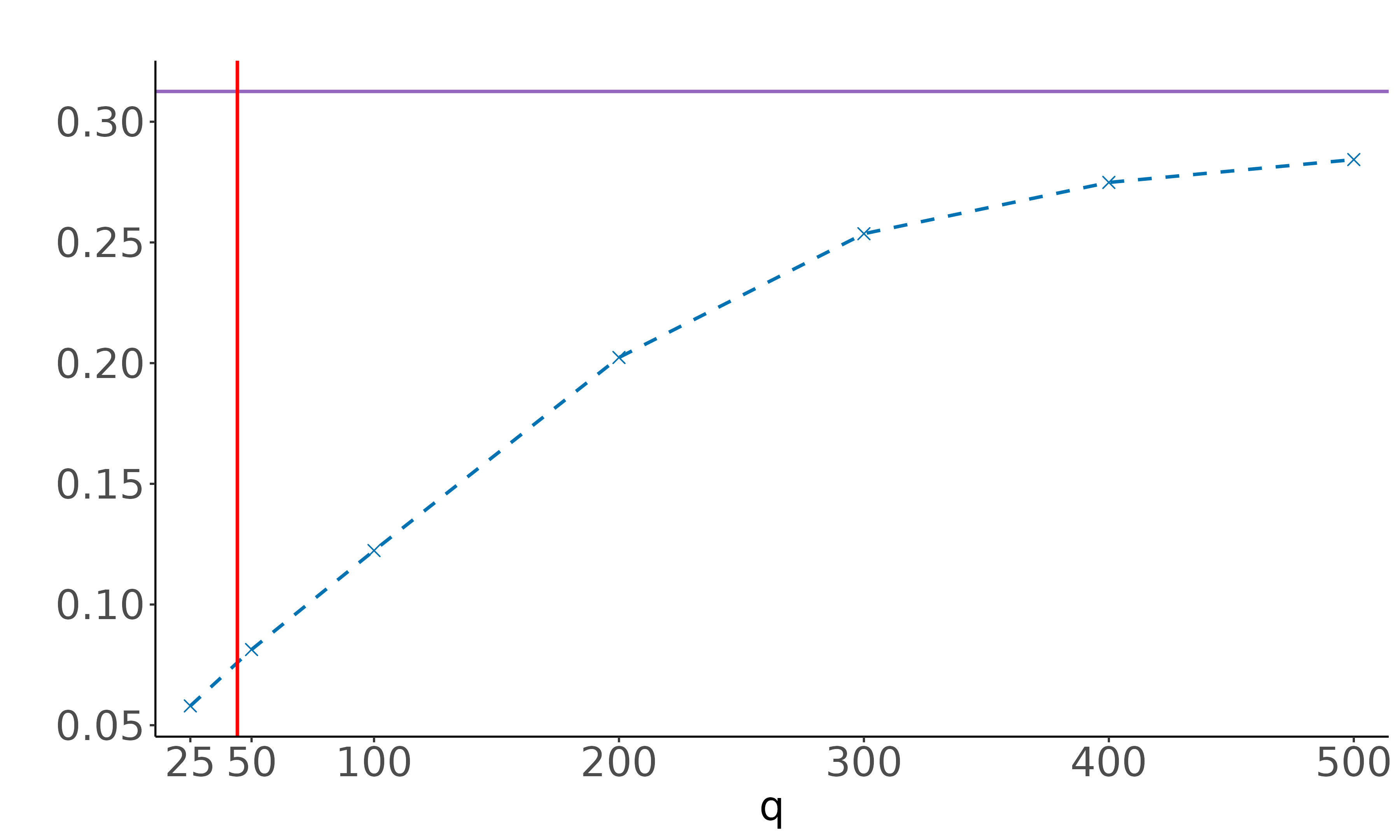}
	\end{subfigure}
      \hspace{-0.5em}

   \includegraphics[width=0.5\textwidth]{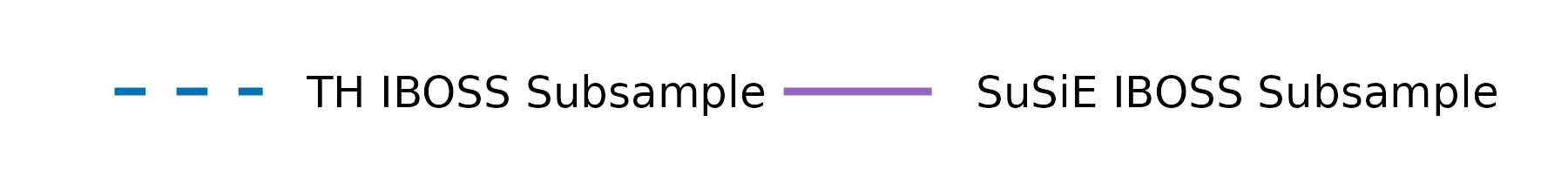} 

	\vspace{2em}
	\centering	\caption{ \it The simulated $r^2$ on test data varying with number of top $q$ variables for TH-IBOSS benchmarked with SuSiE-IBOSS, based on $100$ simulation runs, for linear regression model with $n_{\mathrm{train}}=100{,}000$, $n_{\mathrm{test}}=30{,}000$, $p = 5000$, $c=0.5$, $\rho = 0.95$, $SNR = 0.5$, and $k=5000$. The red vertical line indicates the average number of variables prescreened by SuSiE.}
	\label{Figure: Sensitivity Analysis SIS SuSIe TH} 
\end{figure}

In Figure~\ref{Figure: Sensitivity Analysis SIS SuSIe TH} we see that for `Fixed 12 signal loci' and `Random 50 signal loci', SuSiE prescreens approximately 8 and 44 variables in Stage I (red benchmark line), and achieves $r^2$ of approximately 0.329 and 0.313 on the test data, respectively. In comparison, TH-IBOSS requires retaining about 200 variables in the `Fixed 12 signal loci' setting to perform close to SuSiE-IBOSS, whereas in the `Random 50 signal loci' setting, retaining $500$ variables yields a test \(r^2\) of approximately 0.284, much lower compared to SuSiE-IBOSS (0.313). 

This highlights that, depending on the problem setup (values of $n$, $p$, $B$, and $s$), the SIS implementation of TH-IBOSS may require additional tuning and may not be the best choice.

\section{Additional simulation scenario: moderate covariate correlation}
\begin{figure}[t]
\centering	
\begin{minipage}{0.49\textwidth}
    \centering
    \text{\footnotesize Fixed 12 signal loci }
\end{minipage}
\hfill
\begin{minipage}{0.49\textwidth}
    \centering
    \text{\footnotesize Random 50 signal loci }
\end{minipage}
\vspace{0.0em}
	\begin{subfigure}{.50\textwidth}
		\centering
\includegraphics[width=\linewidth, height=0.3\textheight, keepaspectratio]{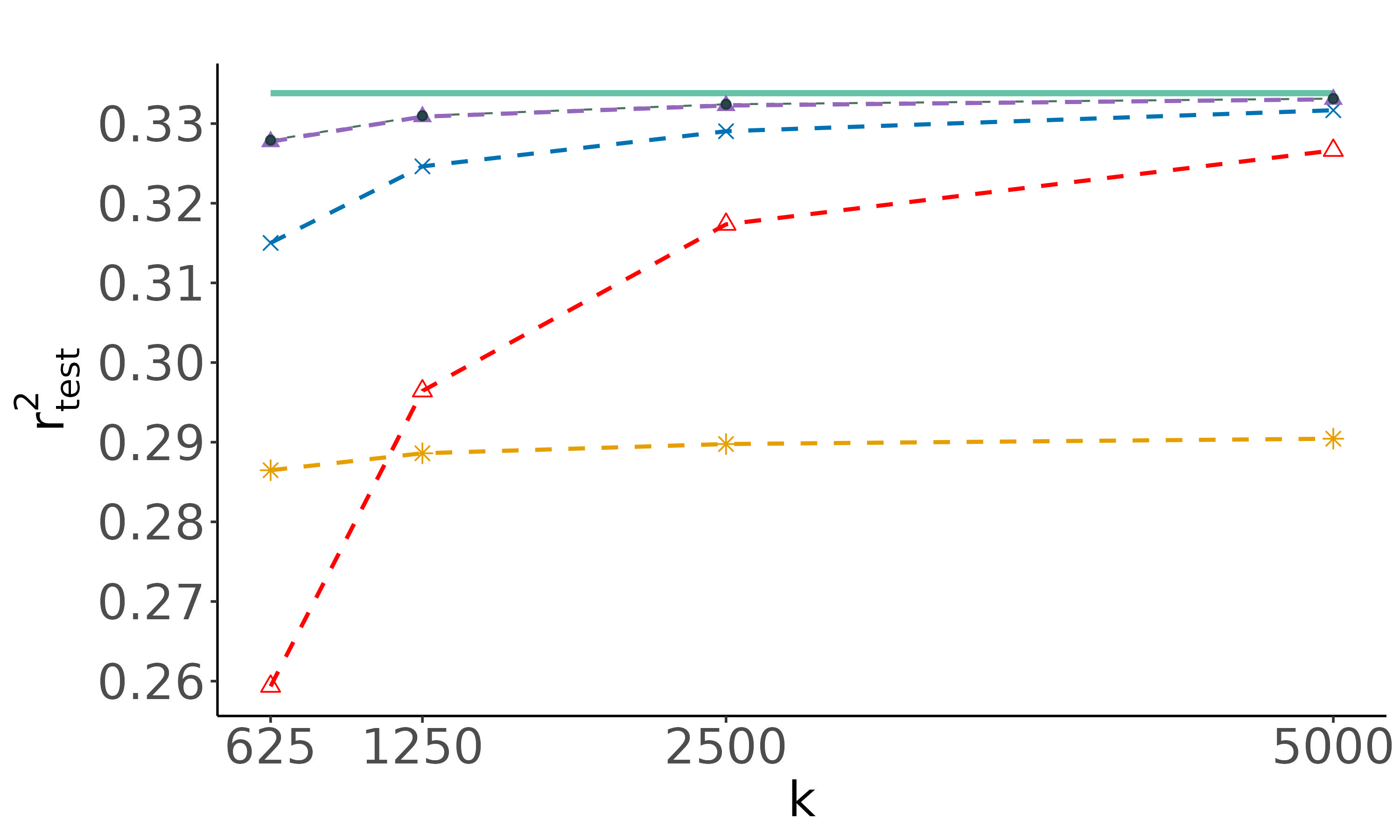}
	\end{subfigure}
      \hspace{-0.5em}
    \begin{subfigure}{.50\textwidth}
		\centering
\includegraphics[width=\linewidth, height=0.3\textheight, keepaspectratio]{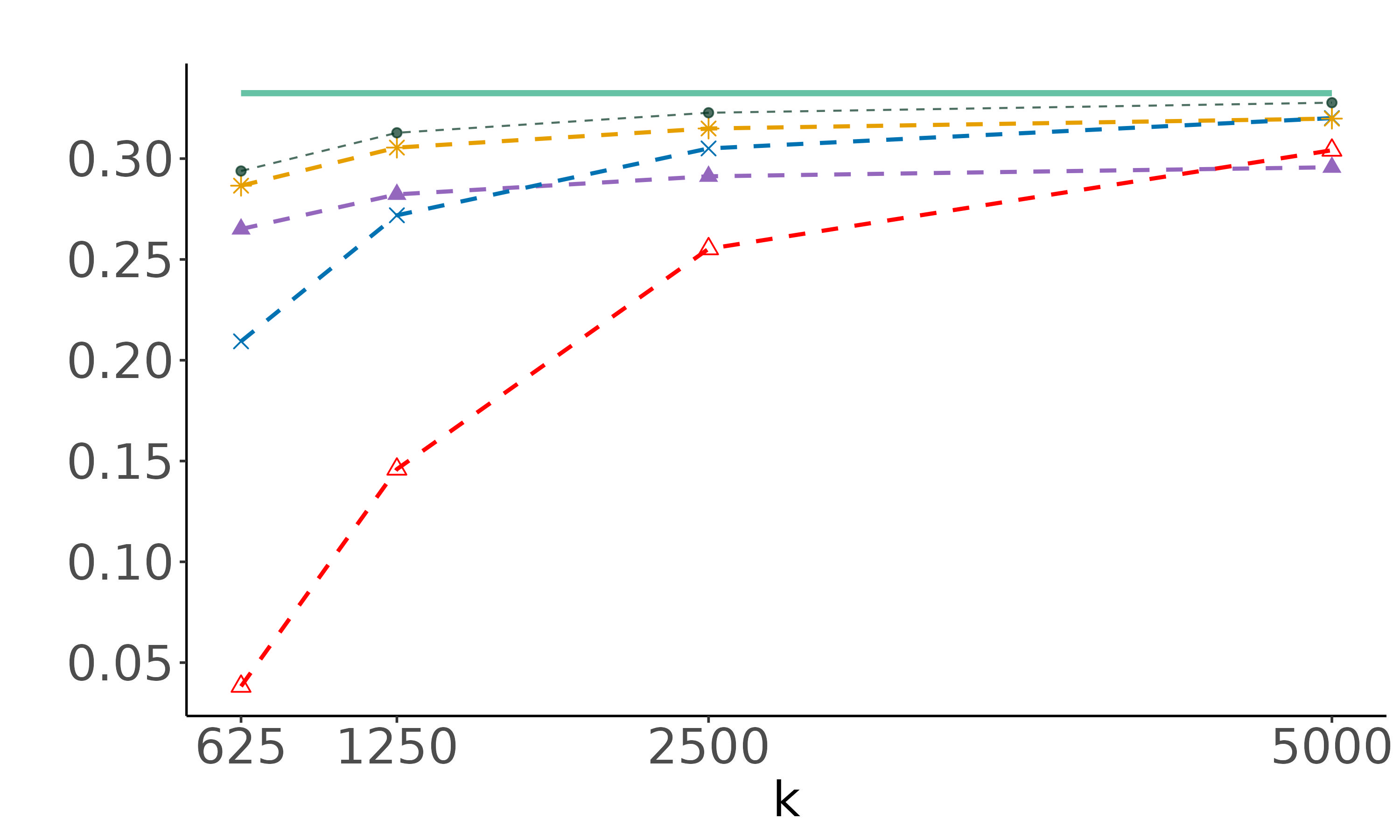}
	\end{subfigure}
{ \it \text{(a) Predictive performance using average $r^2$ on test data}  }	

 \vspace{1.0em}

\centering
\begin{minipage}{0.49\textwidth}
    \centering
    \text{\footnotesize Fixed 12 signal loci }
\end{minipage}
\hfill
\begin{minipage}{0.49\textwidth}
    \centering
    \text{\footnotesize Random 50 signal loci }
\end{minipage}
\vspace{0.0em}
	\begin{subfigure}{.50\textwidth}
		\centering
\includegraphics[width=\linewidth, height=0.3\textheight, keepaspectratio]{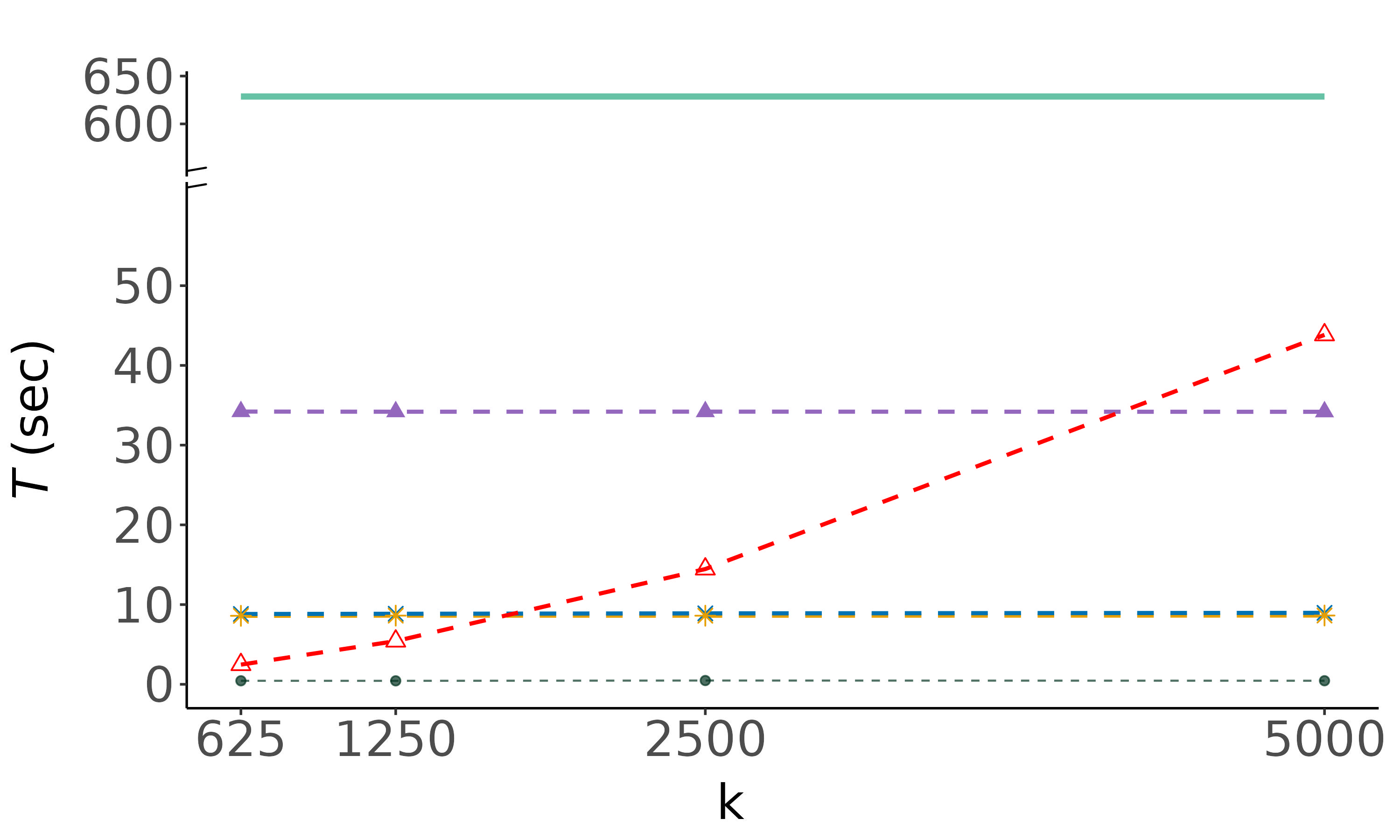}
    \end{subfigure}%
     \hspace{-0.5em}
\begin{subfigure}{.50\textwidth}
		\centering
	\includegraphics[width=\linewidth, height=0.3\textheight, keepaspectratio]{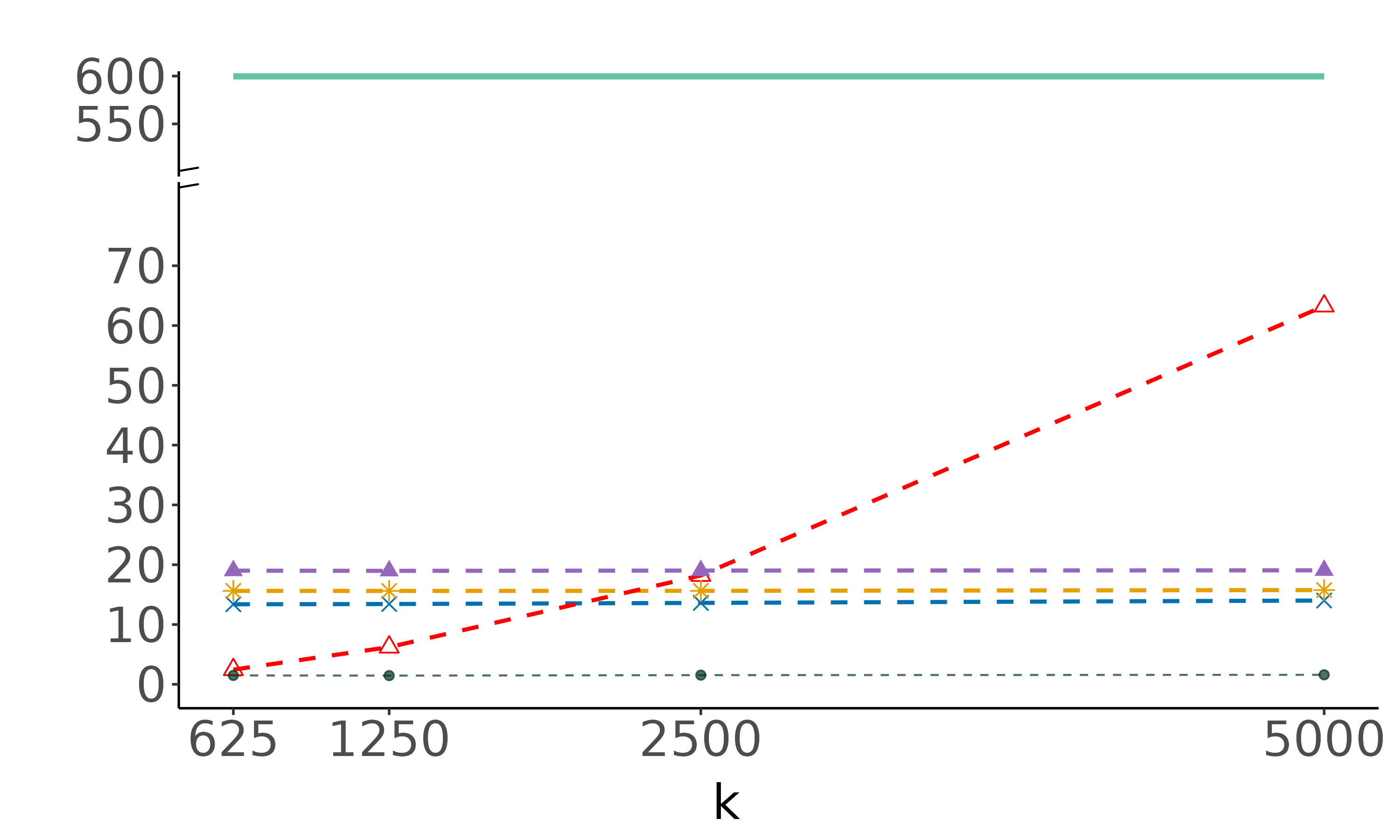}
    \end{subfigure}
      \hspace{-0.5em}
	
  { \it \text{(b) Computational efficiency based on total runtime in seconds}}
\vspace{1.0em} 

   \includegraphics[width=0.95\textwidth]{Plots/by_k_legend_only.png} 

\vspace{0.0em}

\caption{\it Benchmarking scalability of PTS based on $100$ simulation runs
for the linear regression model with $\rho = 0.5$, $SNR = 0.5$,
$n = 10^5$, $p = 5000$, and $c = 0.5$.}
\label{Figure: Sensitivity rho 0.5}
\end{figure}

In this section, we evaluate PTS under moderate covariate correlation, using the same simulation setup as in Section 3 of the main paper but setting $\rho=0.5$. We consider the same linear regression simulation setting with $n=100{,}000$, $n_{\mathrm{test}}=30{,}000$, $c=0.50$, $p=5000$ ($B=5$ and $m=1000$), $\rho=0.50$, $\mathrm{SNR}=0.5$, $\beta_0=1$, under both the `Fixed 12 signal loci' and the `Random signal loci' setting ($s=1\%$). Figure~\ref{Figure: Sensitivity rho 0.5} shows that PTS remains effective under moderate covariate correlation. In the `Fixed 12 signal loci' setting, SuSiE-IBOSS slightly outperforms TH-IBOSS, whereas in the `Random 50 signal loci' setting, CT-IBOSS achieves the highest predictive performance among the non-oracle PTS variants, with TH-IBOSS performing similarly for larger~$k$.

\clearpage

\section{Proof of Proposition 2.1}

 \begin{restatedproposition}[]
\textit{\textbf{Effects of prescreening quality on prediction and informative subsampling}} \\
Let $X_1,X_2,W\sim N(0,1)$  and $\varepsilon\sim N(0,\sigma^2)$ be independent with $\sigma^2>0$. For $\beta_1,\beta_2\neq 0$ and $0<\rho<1$, let
\begin{equation*}
Y=\beta_1X_1+\beta_2X_2+\varepsilon,
\qquad
X_3=\rho X_2+\sqrt{1-\rho^2}W.
\end{equation*}
Thus, $X_1$ and $X_2$ are the relevant predictors, while $X_3$ is a correlated proxy for $X_2$ with no additional effect on $Y$. 

\begin{enumerate}
\item[(i)] \textbf{False-negative prescreening:} Suppose that $0<|\beta_1|<\rho|\beta_2|$, so that selecting the two covariates with the largest absolute population marginal correlations yields  $\widetilde J=\{2,3\}$ instead of
$J'=\{1,2\}$.  Define the minimum squared linear prediction risk as
$R(J)=\inf_{\bm\alpha}
E\bigl[(Y-{\textstyle\sum_{j\in J}}\alpha_jX_j)^2\bigr]$. 
Then,
\begin{equation*}
R(J')=\sigma^2,\qquad
R(\widetilde J)=\sigma^2+\beta_1^2,\qquad
R(\widetilde J)-R(J')=\beta_1^2>0.
\end{equation*}
Thus, omitting $X_1$ increases the minimum squared linear prediction risk by $\beta_1^2$.

\item[(ii)] \textbf{False-positive prescreening:} Suppose one prescreening method selects $J'=\{1,2\}$, whereas another additionally retains the redundant variable~$X_3$ and selects $J''=\{1,2,3\}$. For a common fixed threshold $t>0$, consider an IBOSS-inspired population-tail design defined by
$\xi_{J,t}(\bm{x})=\dfrac{1}{|J|}\sum_{j\in J}f(\bm{x}\mid |X_j|>t)$, where $f(\bm{x})$ denotes the joint population density of $\bm{x}=(X_1,X_2,X_3)^\top$.
Assuming an oracle final model fitted on the true active
variables $\bm{z}=(X_1,X_2)^\top$, let 
$M_J(t)=E_{\xi_{J,t}}
(\bm{z}\bm{z}^\top)$ denote the normalized Fisher information matrix (up to the common factor $1/\sigma^2$). For the A- and D-optimality criteria $\Phi_A(M)=\{\operatorname{tr}(M^{-1})\}^{-1}$ and
$\Phi_D(M)=\log\det(M)$, respectively, we obtain 
\[
\Phi_A(M_{J'}(t))>\Phi_A(M_{J''}(t)), \qquad
\Phi_D(M_{J'}(t))> \Phi_D(M_{J''}(t)).
\] 
Thus, under this population-tail design, including the redundant variable yields a strictly less informative design according to both the A- and D-optimality criteria.
\end{enumerate}

\end{restatedproposition}

\begin{proof}
\noindent \textbf{(i)}
Under $0<|\beta_1|<\rho|\beta_2|$, screening by absolute population marginal correlation retains $\widetilde J=\{2,3\}$, since $|\operatorname{Corr}(X_2,Y)|>|\operatorname{Corr}(X_3,Y)|>|\operatorname{Corr}(X_1,Y)|$. This follows from $\operatorname{Cov}(X_1,Y)=\beta_1$, $\operatorname{Cov}(X_2,Y)=\beta_2$, $\operatorname{Cov}(X_3,Y)=\rho\beta_2$, and unit variances of all three covariates.

\noindent For $J'=\{1,2\}$, $Y-\alpha_1X_1-\alpha_2X_2=(\beta_1-\alpha_1)X_1+(\beta_2-\alpha_2)X_2+\varepsilon$. Since $X_1$, $X_2$, and $\varepsilon$ are independent and centered, $E\bigl[(Y-\alpha_1X_1-\alpha_2X_2)^2\bigr]=(\beta_1-\alpha_1)^2+(\beta_2-\alpha_2)^2+\sigma^2$, which is minimized at $\alpha_1=\beta_1$ and $\alpha_2=\beta_2$. Hence $R(J')=\sigma^2$.

\noindent For $\widetilde J=\{2,3\}$, $Y-\alpha_2X_2-\alpha_3X_3=\beta_1X_1+(\beta_2-\alpha_2)X_2-\alpha_3X_3+\varepsilon$. Since $X_1$ is independent of $(X_2,X_3,\varepsilon)$ and centered, $E\bigl[(Y-\alpha_2X_2-\alpha_3X_3)^2\bigr]=\beta_1^2+\sigma^2+E\bigl[
\{(\beta_2-\alpha_2)X_2-\alpha_3X_3\}^2
\bigr]$. The final term is nonnegative and equals zero at $\alpha_2=\beta_2$ and $\alpha_3=0$. Therefore $R(\widetilde J)=\sigma^2+\beta_1^2$.\\

\noindent This shows that omitting $X_1$ increases the minimum squared linear prediction risk by $\beta_1^2$.

\noindent \textbf{(ii)} We compare the prescreening sets $J'=\{1,2\}$
and $J''=\{1,2,3\}$ under the population-tail
design at a common fixed threshold $t>0$. 
The normalized information matrix can be written directly as
\[
M_J(t)
=
E_{\xi_{J,t}}(\bm{z}\bm{z}^\top)
=
\begin{pmatrix}
E_{\xi_{J,t}}(X_1^2) &
E_{\xi_{J,t}}(X_1X_2)\\
E_{\xi_{J,t}}(X_1X_2) &
E_{\xi_{J,t}}(X_2^2)
\end{pmatrix}.
\]
Let \(U\sim N(0,1)\) and, for \(t>0\), define
\[
v_t=\mathbb{E}(U^2\mid |U|>t)
=1+t\frac{f_N(t)}{1-F_N(t)}>1,
\]
where \(f_N\) and \(F_N\) denote the standard normal density and cumulative distribution function, respectively.

\medskip
\noindent\textbf{Case 1: $J'=\{1,2\}$}
\[
\xi_{J',t}(\bm{x})
=
\frac12f(\bm{x}\mid |X_1|>t)
+
\frac12f(\bm{x}\mid |X_2|>t),
\]
and hence
\[
M_{J'}(t)
=
\frac12E(\bm{z}\bm{z}^\top\mid |X_1|>t)
+
\frac12E(\bm{z}\bm{z}^\top\mid |X_2|>t).
\]
For the $X_1$-tail contribution, $E(X_1^2\mid |X_1|>t)=v_t$, $E(X_2^2\mid |X_1|>t)=1$ by independence, and $E(X_1X_2\mid |X_1|>t)=0$. Therefore
\[
E(\bm{z}\bm{z}^\top\mid |X_1|>t)
=
\begin{pmatrix}
v_t&0\\
0&1
\end{pmatrix}.
\]
By symmetry, interchanging $X_1$ and $X_2$ gives
\[
E(\bm{z}\bm{z}^\top\mid |X_2|>t)
=
\begin{pmatrix}
1&0\\
0&v_t
\end{pmatrix}.
\]
Thus
\[
M_{J'}(t)
=
\frac12
\left\{
\begin{pmatrix}
v_t&0\\
0&1
\end{pmatrix}
+
\begin{pmatrix}
1&0\\
0&v_t
\end{pmatrix}
\right\}
=
\frac{v_t+1}{2}I_2.
\]

\medskip
\noindent\textbf{Case 2: $J''=\{1,2,3\}$.}
For $J''=\{1,2,3\}$,
\[
\xi_{J'',t}(\bm{x})
=
\frac13f(\bm{x}\mid |X_1|>t)
+
\frac13f(\bm{x}\mid |X_2|>t)
+
\frac13f(\bm{x}\mid |X_3|>t),
\]
and therefore
\[
M_{J''}(t)
=
\frac13E(\bm{z}\bm{z}^\top\mid |X_1|>t)
+
\frac13E(\bm{z}\bm{z}^\top\mid |X_2|>t)
+
\frac13E(\bm{z}\bm{z}^\top\mid |X_3|>t).
\]
The first two conditional matrices are those obtained above. 

\noindent It remains to calculate the $X_3$-tail contribution. Since $X_1$ is independent of $(X_2,X_3)$, $E(X_1^2\mid |X_3|>t)=1$ and $E(X_1X_2\mid |X_3|>t)=0$. Since $(X_2,X_3)$ are jointly standard normal with $\operatorname{Corr}(X_2,X_3)=\rho$, the conditional distribution of $X_2$ given $X_3=x_3$ is $ X_2\mid X_3=x_3 \sim N(\rho x_3,1-\rho^2).$
Hence
\[
E(X_2^2\mid X_3=x_3)
=
\operatorname{Var}(X_2\mid X_3=x_3)
+
\{E(X_2\mid X_3=x_3)\}^2
=
1-\rho^2+\rho^2x_3^2.
\]
Now let $A_t=\{|X_3|>t\}$. Since $A_t\in\sigma(X_3)$, the conditional law of iterated expectation gives
\[
\begin{aligned}
E(X_2^2\mid A_t)
&=
E\!\left[E(X_2^2\mid X_3)\mid A_t\right]\\
&=
1-\rho^2+\rho^2E(X_3^2\mid |X_3|>t)\\
&=
1-\rho^2+\rho^2v_t\\
&=
1+\rho^2(v_t-1).
\end{aligned}
\]
Thus
\[
E(\bm{z}\bm{z}^\top\mid |X_3|>t)
=
\begin{pmatrix}
1&0\\
0&\rho^2v_t+1-\rho^2
\end{pmatrix}.
\]
Combining the three contributions,
\[
M_{J''}(t)
=
\frac13
\left\{
\begin{pmatrix}
v_t&0\\
0&1
\end{pmatrix}
+
\begin{pmatrix}
1&0\\
0&v_t
\end{pmatrix}
+
\begin{pmatrix}
1&0\\
0&\rho^2v_t+1-\rho^2
\end{pmatrix}
\right\},
\]
so
\[
M_{J''}(t)
=
\begin{pmatrix}
(v_t+2)/3&0\\
0&\{(1+\rho^2)v_t+2-\rho^2\}/3
\end{pmatrix}.
\]

\medskip 
\medskip 

\noindent For A-optimality,  \[\Phi_A(M_{J'}(t))=(v_t+1)/4 \text{ and } \Phi_A(M_{J''}(t))=\frac{(v_t+2)\{(1+\rho^2)v_t+2-\rho^2\}}{3\{(2+\rho^2)v_t+4-\rho^2\}}.
\]
Therefore for $d_t=v_t-1>0$, we obtain
\[
\Phi_A(M_{J'}(t))-\Phi_A(M_{J''}(t))=\frac{d_t\{(2-\rho^2)d_t+6(1-\rho^2)\}}{12\{(2+\rho^2)v_t+4-\rho^2\}}>0,
\]
since $v_t>1$ and $0<\rho<1$.

\noindent For D-optimality, the determinants are $\det M_{J'}(t)=(v_t+1)^2/4$ and $\det M_{J''}(t)=(v_t+2)\{(1+\rho^2)v_t+2-\rho^2\}/9$. Therefore, 
\[
\Phi_D(M_{J'}(t))-\Phi_D(M_{J''}(t))=\log\left(1+\frac{12(1-\rho^2)d_t+(5-4\rho^2)d_t^2}{4(3+d_t)\{3+(1+\rho^2)d_t\}}\right)>0,
\]
since $d_t>0$ and $0<\rho<1$, so the argument of the logarithm is strictly greater than one. 
\end{proof}



\end{document}